\documentclass[prd,aps,
superscriptaddress]{revtex4}

\usepackage{xcolor}
\usepackage{graphicx}
\usepackage[caption=false]{subfig}

\usepackage{amsmath}
\usepackage{booktabs}
\usepackage{multirow}
\usepackage{rotating}

\usepackage{bm}        
\usepackage{txfonts}   
\usepackage{hyperref}  

\date{\today}
\begin{document}
%

%
\title{
Probing the Moderate-to-Large-\(x\) Gluon Helicity with \(J/\psi\) Production at \(\sqrt{s}=27~\mathrm{GeV}\)
%
%
}
\author{Shubham Sharma}
\email[]{s.sharma.hep@gmail.com}
\affiliation{Moscow Institute of Physics and Technology, Dolgoprudny 141700, Russia}

\author{Alexey Aparin}
\email[]{aparin@jinr.ru}
\affiliation{Moscow Institute of Physics and Technology, Dolgoprudny 141700, Russia}
\affiliation{Joint Institute for Nuclear Research, Dubna 141980, Russia}
\affiliation{Institute for Nuclear Physics, Almaty, 050032, Kazakhstan}

\author{Satyajit Puhan}
\email[]{puhansatyajit@gmail.com}
\affiliation{Institute of Physics, Academia Sinica, Taipei 11529, Taiwan}

\author{Narinder Kumar}
\email[]{narinderhep@gmail.com}
\affiliation{Computational Theoretical High Energy Physics Lab, Department of Physics, Doaba College, Jalandhar 144004, India}

\author{Harleen Dahiya}
\email[]{dahiyah@nitj.ac.in}
\affiliation{Computational High Energy Physics Lab, Department of Physics, Dr. B.R. Ambedkar National Institute of Technology, Jalandhar 144008, India}

\date{\today}
%
%
%
\begin{abstract}
We investigate the longitudinal double-spin asymmetry, $A_{LL}$, for color-singlet $J/\psi$ production in longitudinally polarized proton-proton collisions at $\sqrt{s}\simeq27~\mathrm{GeV}$, relevant to the Spin Physics Detector (SPD) at the Nuclotron-based Ion Collider fAcility (NICA). At these relatively low collision energies, $J/\psi$ production through gluon fusion probes gluon momentum fractions in the moderate-to-large-$x$ region, with increasingly asymmetric partonic configurations at forward rapidity. We consider the leading-order color-singlet subprocess $g g\to J/\psi+g$ within the nonrelativistic quantum chromodynamics (QCD) framework and calculate the polarized and unpolarized cross sections using the corresponding analytic partonic hard-scattering coefficients. Since the same color-singlet long-distance matrix element enters the polarized and unpolarized cross sections, it cancels out in the ratio defining $A_{LL}$ at this order. We evaluate the transverse-momentum and rapidity dependence of $A_{LL}$ after convolution with polarized and unpolarized gluon distributions. The sensitivity of the asymmetry to different combinations of unpolarized and polarized PDF sets is examined, including both leading order (LO) and next-to-leading order (NLO) PDF inputs, while the hard-scattering coefficient is kept at leading order throughout the research. Factorization- and renormalization-scale variations are also studied, and PDF and scale uncertainties are combined to quantify the theoretical uncertainty. The resulting asymmetry is positive and increases with transverse momentum over the kinematically stable region, while its magnitude and uncertainty become increasingly sensitive to the polarized gluon distribution at moderate and large $x$. We find substantial differences among contemporary PDF combinations at moderate and large transverse momentum. The uncertainty analysis shows that the relative importance of PDF and scale uncertainties is kinematics dependent: NLO PDF uncertainties are generally reduced relative to LO in central regions, but this reduction is not uniform and both PDF and scale uncertainties can become sizable at large transverse momentum and forward rapidity. These results indicate that differential measurements of $A_{LL}$ at SPD could provide complementary constraints on the gluon helicity distribution in a kinematic region that is presently less constrained. A complete NLO calculation of the hard-scattering coefficient will be required for precision predictions and for a quantitative assessment of higher-order effects. 
%

\end{abstract}
%
\maketitle
%
%
\section{Introduction}

Understanding the origin of the proton spin remains a central challenge in QCD. Following the discovery by the European Muon Collaboration (EMC) that the quark helicity contribution accounts for only a fraction of the proton spin~\cite{EuropeanMuon:1987isl,EuropeanMuon:1989yki}, considerable experimental and theoretical efforts have been devoted to determining the contributions from gluon polarization and partonic orbital angular momentum~\cite{Jaffe:1989jz,Ji:1996ek,Leader:2013jra}. Global analyses of polarized deep-inelastic scattering and polarized proton-proton data have established evidence for a nonzero gluon helicity distribution, $\Delta g(x)$, in the intermediate-$x$ region~\cite{deFlorian:2014yva,Nocera:2014gqa,Ethier:2017zbq}. Nevertheless, substantial uncertainties remain, particularly at moderate and large momentum fractions, where experimental constraints on $\Delta g(x)$ are still limited~\cite{Ethier:2017zbq}.

Measurements at the Relativistic Heavy Ion Collider (RHIC) have provided important constraints on the gluon helicity distribution through inclusive jet and hadron production in longitudinally polarized proton-proton collisions at $\sqrt{s}=200$ and $510~\mathrm{GeV}$~\cite{STAR:2014wox,STAR:2021mqa,PHENIX:2014gbf}. These measurements predominantly constrain $\Delta g(x)$ at relatively small and intermediate momentum fractions. Future measurements at the Electron-Ion Collider (EIC) are expected to extend the available constraints toward even smaller values of $x$~\cite{Accardi:2012qut}. Complementary information can therefore be obtained from experiments operating at substantially lower collision energies, where the production of hard probes requires partons carrying larger fractions of the proton momentum. In this respect, the moderate center-of-mass energies available at the Nuclotron-based Ion Collider fAcility (NICA) provide access to a kinematic regime in which the gluon helicity distribution at intermediate and large $x$ can be investigated~\cite{Arbuzov:2020cqg}.

The SPD at NICA is designed to investigate the spin structure of protons and deuterons using longitudinally and transversely polarized beams, with proton-proton collisions envisaged at center-of-mass energies up to approximately $\sqrt{s}=27~\mathrm{GeV}$~\cite{Guskov:2023dlj,Arbuzov:2020cqg}. Among the processes considered in the SPD physics program, charmonium production is of particular interest because of its sensitivity to gluon-initiated subprocesses and, consequently, to the gluon structure of the proton~\cite{Guskov:2023dlj,Arbuzov:2020cqg}. In particular, $J/\psi$ production at SPD energies has recently been investigated in detail, including its transverse-momentum and rapidity dependence and the contributions of different quarkonium production mechanisms~\cite{Sharma:2025mnu}. The longitudinal double-spin asymmetry, $A_{LL}$, in inclusive $J/\psi$ production consequently provides a means of studying the polarization dependence of gluon-initiated production and, through the polarized gluon parton distribution, of probing the gluon helicity structure of the proton. Earlier theoretical studies have demonstrated the sensitivity of $J/\psi$ double-spin asymmetries to the polarized gluon distribution at fixed-target energies~\cite{Teryaev:1996sr}, while more recent phenomenological studies have investigated the observable at RHIC energies within the nonrelativistic QCD framework~\cite{Feng:2018cai}.

The sensitivity of $J/\psi$ production to gluon polarization has also been investigated experimentally at RHIC. The PHENIX Collaboration measured the double-helicity asymmetry $A_{LL}^{J/\psi}$ in inclusive $J/\psi$ production in longitudinally polarized proton-proton collisions at $\sqrt{s}=510~\mathrm{GeV}$~\cite{PHENIX:2016gzj}. The measurement demonstrated the feasibility of using this observable to constrain the gluon helicity distribution, with the relevant production process receiving substantial contributions from gluon-gluon scattering. On the theoretical side, Feng and Zhang calculated the $J/\psi$ double-longitudinal-spin asymmetry at next-to-leading order in QCD and studied its dependence on different sets of long-distance matrix elements~\cite{Feng:2018cai}. These studies provide an established high-energy basis for investigating $A_{LL}^{J/\psi}$ as a probe of gluon polarization in polarized proton-proton collisions.

The substantially lower collision energies of SPD/NICA shift the relevant partonic kinematics toward larger gluon momentum fractions. At $\sqrt{s}\approx27~\mathrm{GeV}$, $J/\psi$ production at central rapidity probes gluons with momentum fractions of order $x>0.1$, while increasing the rapidity of the produced $J/\psi$ can extend the kinematic reach toward extremely-low or -high momentum fractions. The SPD/NICA energy range therefore provides a complementary opportunity to investigate the gluon helicity distribution in a region that is not well constrained by existing high-energy measurements. This motivates a dedicated study of $A_{LL}^{J/\psi}$ in the less explored energy regime.

In this work, we investigate the longitudinal double-spin asymmetry $A_{LL}$ for color-singlet $J/\psi$ production in longitudinally polarized proton-proton collisions at the center-of-mass energy relevant for the SPD at NICA, focusing on $\sqrt{s}=27~\mathrm{GeV}$. Building on our previous study of inclusive unpolarized $J/\psi$ production at SPD energies, where the dependence on the gluon distributions, the contributions of different quarkonium production mechanisms, and the corresponding differential cross sections were investigated~\cite{Sharma:2025mnu}, we turn here to the spin-dependent aspect of the same production process. In particular, we calculate $A_{LL}$ for the color-singlet contribution through the leading-order subprocess $g g\to J/\psi+g$, and investigate its dependence on the transverse momentum $p_T$ and rapidity $y_{J/\psi}$. The polarized and unpolarized production cross sections are obtained by convoluting the corresponding partonic hard-scattering coefficients with polarized and unpolarized gluon distributions. The calculation employs the CT18LO~\cite{Yan:2022pzl}, CT18NLO~\cite{Hou:2019qau}, and NNPDF3.1 LO and NLO~\cite{NNPDF:2017mvq} unpolarized PDF sets in combination with the NNPDFpol2.0 LO and NLO polarized gluon distributions~\cite{Cruz-Martinez:2025ahf}, accessed through the LHAPDF framework~\cite{Buckley:2014ana}. We study the rapidity and transverse-momentum dependence of $A_{LL}$ and examine its sensitivity to the choice and perturbative order of the input PDFs. In addition, we quantify the theoretical uncertainties associated with the PDF parametrizations and with variations of the factorization and renormalization scales. Throughout this work, the hard-scattering coefficient is evaluated at leading order; the calculations employing NLO PDF sets are used only to assess the sensitivity to the perturbative order of the parton distributions and do not constitute complete NLO predictions. By isolating the color-singlet contribution, this study provides a well-defined theoretical baseline for $A_{LL}$ in $J/\psi$ production at SPD/NICA and assesses the sensitivity of this observable to the polarized gluon distribution in the moderate-$x$ region relevant to the experiment.
%
\section{Theoretical Framework}
\label{sec:theoretical_framework}
\subsection{Kinematics of color-singlet \(J/\psi\) production}
\label{subsec:CS_kinematics}

We restrict the present analysis to the color-singlet (CS) mechanism for
prompt $J/\psi$ production. At LO in the collinear
factorization framework, the relevant partonic subprocess is
\begin{equation}
g(k_1)+g(k_2)
\rightarrow
c\bar c[{}^3S_1^{[1]}](p)+g(k_3)
\rightarrow
J/\psi(p)+g(k_3).
\label{eq:CS_2to2_process}
\end{equation}
The final-state gluon provides the recoil required to produce a
$J/\psi$ with nonzero transverse momentum. Consequently, the CS
contribution considered here is a $2\to2$ process, and the transverse
momentum of the $J/\psi$ is balanced by that of the recoiling gluon,
\begin{equation}
\boldsymbol{p}_T+\boldsymbol{k}_{T3}=0,
\qquad
|\boldsymbol{k}_{T3}|=p_T.
\label{eq:transverse_momentum_balance}
\end{equation}
The independent final-state kinematic variables may therefore be taken
as the $J/\psi$ transverse momentum $p_T$ and rapidity $y_{J/\psi}$, together with the rapidity of the
recoiling gluon $y_g$ .

Let $P_1$ and $P_2$ denote the four-momenta of the incoming protons,
with
\begin{equation}
s=(P_1+P_2)^2,
\end{equation}
and let the momenta of the incoming partons be
\begin{equation}
k_1=x_1P_1,
\qquad
k_2=x_2P_2.
\end{equation}
Following the convention used in the short-distance coefficient (SDC) of
Ref.~\cite{Klasen:2003zn}, we denote the invariant mass of the
produced $c\bar c$ pair by
\begin{equation}
M=2m_c,
\label{eq:M_CS_definition}
\end{equation}
where $m_c$ is the charm-quark mass. The corresponding transverse mass
is
\begin{equation}
m_T=\sqrt{M^2+p_T^2}.
\label{eq:mT_CS}
\end{equation}

Energy and longitudinal momentum conservation for the $2\to2$
subprocess then give the incoming partonic momentum fractions as
\begin{equation}
x_1=
\frac{
m_T e^{y_{J/\psi}}
+p_T e^{y_g}
}{
\sqrt{s}
},
\qquad
x_2=
\frac{
m_T e^{-y_{J/\psi}}
+p_T e^{-y_g}
}{
\sqrt{s}
}.
\label{eq:x1x2_CS}
\end{equation}
The physical phase space is restricted by
\begin{equation}
0<x_1\leq1,
\qquad
0<x_2\leq1.
\label{xcons}
\end{equation}

For fixed $p_T$ and $y_{J/\psi}$, these conditions determine the
allowed range of the recoil-gluon rapidity, denoted by
$y_g^{\rm min}\leq y_g\leq y_g^{\rm max}$.

Equation~\eqref{eq:x1x2_CS} also makes explicit that the gluon momentum
fractions probed in the CS $2\to2$ process depend on the complete
final-state configuration. In particular,
\begin{equation}
x_1\geq
\frac{m_T}{\sqrt{s}}e^{y_{J/\psi}},
\qquad
x_2\geq
\frac{m_T}{\sqrt{s}}e^{-y_{J/\psi}},
\end{equation}
with the additional contribution determined by the recoil gluon.
Thus, unlike a $2\to1$ kinematic configuration, the momentum fractions
cannot be specified solely in terms of the observed $J/\psi$
kinematics.

The partonic Mandelstam variables for the subprocess in
Eq.~\eqref{eq:CS_2to2_process} are
\begin{align}
\hat{s} &= x_1x_2s,
\label{eq:hat_s_CS}
\\
\hat{t} &= M^2-x_1\sqrt{s}\,m_T e^{-y_{J/\psi}},
\label{eq:hat_t_CS}
\\
\hat{u} &= M^2-x_2\sqrt{s}\,m_T e^{y_{J/\psi}},
\label{eq:hat_u_CS}
\end{align}
and satisfy
\begin{equation}
\hat{s}+\hat{t}+\hat{u}=M^2.
\label{eq:mandelstam_relation_CS}
\end{equation}
These relations provide the kinematic basis for the convolution of the
partonic hard-scattering coefficients with the gluon distributions in
the calculation of the longitudinal double-spin asymmetry.

\subsection{Longitudinal double-spin asymmetry}
\label{subsec:ALL}

The longitudinal double-spin asymmetry for inclusive $J/\psi$
production in longitudinally polarized proton-proton collisions is
defined as~\cite{Feng:2018cai,Klasen:2003zn}
\begin{equation}
A_{LL}
=
\frac{d\Delta\sigma}{d\sigma},
\label{eq:ALL_def}
\end{equation}
where $d\sigma$ and $d\Delta\sigma$ denote the unpolarized and
helicity-dependent polarized differential cross sections,
respectively. In terms of the helicity-dependent cross sections
$d\sigma^{\xi_A\xi_B}$, where $\xi_A,\xi_B=\pm1$ denote the helicities
of the two incoming protons, they are given by
\begin{align}
d\sigma
&=
\frac{1}{4}
\sum_{\xi_A,\xi_B=\pm1}
d\sigma^{\xi_A\xi_B},
\label{eq:unpolarized_helicity_average}
\\
d\Delta\sigma
&=
\frac{1}{4}
\sum_{\xi_A,\xi_B=\pm1}
\xi_A\xi_B\,
d\sigma^{\xi_A\xi_B}.
\label{eq:delta_sigma_helicity}
\end{align}

Within collinear QCD factorization, the polarized and unpolarized
cross sections are obtained by convoluting the corresponding
partonic hard-scattering coefficients with the parton distribution
functions (PDFs). For the gluon-gluon initiated CS contribution,
these expressions reduce to
\begin{align}
d\sigma_{\rm CS}
&=
\int dx_1\,dx_2\,
g(x_1,\mu_F)\,
g(x_2,\mu_F)\,
d\hat{\sigma}_{\rm CS},
\label{eq:unpolarized_CS_factorization}
\\
d\Delta\sigma_{\rm CS}
&=
\int dx_1\,dx_2\,
\Delta g(x_1,\mu_F)\,
\Delta g(x_2,\mu_F)\,
d\Delta\hat{\sigma}_{\rm CS},
\label{eq:polarized_CS_factorization}
\end{align}
where $g(x,\mu_F)$ and $\Delta g(x,\mu_F)$ denote the unpolarized and
helicity-dependent gluon distributions, respectively, and $\mu_F$ is
the factorization scale.

For a given partonic channel, the polarized and unpolarized hard parts
are related point by point in phase space through the partonic
longitudinal analyzing power,
\begin{equation}
d\Delta\hat{\sigma}_{\rm CS}
=
\hat a_{LL}^{\rm CS}(\hat{s},\hat{t},\hat{u})\,
d\hat{\sigma}_{\rm CS},
\label{eq:delta_sigma_from_all}
\end{equation}
where
\begin{equation}
\hat a_{LL}^{\rm CS}
=
\frac{
d\Delta\hat{\sigma}_{\rm CS}
}{
d\hat{\sigma}_{\rm CS}
}.
\label{eq:partonic_all_CS_general}
\end{equation}
Since $\hat a_{LL}^{\rm CS}$ depends on the partonic kinematics, it
must remain inside the convolution integrals. In particular, the
hadronic asymmetry cannot in general be obtained by taking a ratio of
separately integrated partonic analyzing powers.

Because the present analysis contains only a single NRQCD channel,
the corresponding color-singlet long-distance matrix element (LDME) $\langle {\cal O}^{J/\psi}({}^3S_1^{[1]})\rangle$ is a common multiplicative factor in the polarized and unpolarized cross sections and therefore cancels in their ratio \cite{Butenschoen:2011yh,Gong:2012ug,Bodwin:2014gia,Zhang:2014ybe,Sun:2015pia}. The resulting $A_{LL}^{\rm CS}$ is consequently
independent of the overall normalization of the color-singlet LDME .

\subsection{Color-singlet contribution to $J/\psi$ spin asymmetry}
\label{subsec:CS_ALL}

For the CS subprocess in Eq.~\eqref{eq:CS_2to2_process}, the
longitudinal double-spin asymmetry is therefore given by
\begin{equation}
A_{LL}^{\rm CS}
=
\frac{
\displaystyle
\int dx_1\,dx_2\,
\Delta g(x_1,\mu_F)
\Delta g(x_2,\mu_F)
d\Delta\hat{\sigma}_{\rm CS}
}{
\displaystyle
\int dx_1\,dx_2\,
g(x_1,\mu_F)
g(x_2,\mu_F)
d\hat{\sigma}_{\rm CS}
}.
\label{eq:ALL_CS}
\end{equation}
Using Eq.~\eqref{eq:delta_sigma_from_all}, this may equivalently be
written as
\begin{equation}
A_{LL}^{\rm CS}
=
\frac{
\displaystyle
\int dx_1\,dx_2\,
\Delta g(x_1,\mu_F)
\Delta g(x_2,\mu_F)
\hat a_{LL}^{\rm CS}(\hat{s},\hat{t},\hat{u})
d\hat{\sigma}_{\rm CS}
}{
\displaystyle
\int dx_1\,dx_2\,
g(x_1,\mu_F)
g(x_2,\mu_F)
d\hat{\sigma}_{\rm CS}
}.
\label{eq:ALL_CS_all}
\end{equation}
This form makes explicit that both the polarized gluon distributions
and the partonic analyzing power contribute to the hadronic
asymmetry.

The SDC for the CS subprocess are taken from
the analytic helicity-dependent calculation of
Ref.~\cite{Klasen:2003zn}, in particular from Eq.~(A.17) of that
reference. With the convention $M=2m_c$, the unpolarized CS partonic
cross section is written as
\begin{equation}
\frac{d\hat{\sigma}_{\rm CS}}{d\hat{t}}
=
\mathcal{N}
\left[
\hat{t}^{2}\hat{u}^{2}
+\hat{s}\hat{t}\hat{u}(\hat{t}+\hat{u})
+\hat{s}^{2}
\left(
\hat{t}^{2}
+\hat{t}\hat{u}
+\hat{u}^{2}
\right)
\right],
\label{eq:CS_unpolarized}
\end{equation}
where
\begin{equation}
\mathcal{N}
=
\frac{
5g_s^6
\langle {\cal O}^{J/\psi}({}^3S_1^{[1]})\rangle M
}{
216\pi
\hat{s}^{2}
(\hat{s}+\hat{t})^{2}
(\hat{s}+\hat{u})^{2}
(\hat{t}+\hat{u})^{2}
}.
\label{eq:CS_normalization}
\end{equation}
The corresponding polarized partonic cross section is
\begin{equation}
\frac{d\Delta\hat{\sigma}_{\rm CS}}{d\hat{t}}
=
\mathcal{N}
\left[
-\hat{t}^{2}\hat{u}^{2}
-\hat{s}\hat{t}\hat{u}(\hat{t}+\hat{u})
+\hat{s}^{2}\hat{t}\hat{u}
\right].
\label{eq:CS_polarized}
\end{equation}
The strong coupling is evaluated at the renormalization scale,
\begin{equation}
g_s^2=4\pi\alpha_s(\mu_R).
\label{eq:gs_alpha_s}
\end{equation}

Taking the ratio of Eqs.~\eqref{eq:CS_polarized} and
\eqref{eq:CS_unpolarized}, the partonic analyzing power becomes
\begin{equation}
\hat a_{LL}^{\rm CS}
=
\frac{
-\hat{t}^{2}\hat{u}^{2}
-\hat{s}\hat{t}\hat{u}(\hat{t}+\hat{u})
+\hat{s}^{2}\hat{t}\hat{u}
}{
\hat{t}^{2}\hat{u}^{2}
+\hat{s}\hat{t}\hat{u}(\hat{t}+\hat{u})
+\hat{s}^{2}
\left(
\hat{t}^{2}
+\hat{t}\hat{u}
+\hat{u}^{2}
\right)
}.
\label{eq:aLL_CS}
\end{equation}
All common factors in $\mathcal{N}$, including the color-singlet LDME,
the strong coupling, and the overall normalization, cancel in this
ratio. Hence, $\hat a_{LL}^{\rm CS}$ is determined entirely by the
partonic kinematics.

\subsection{Hadronic cross sections and phase-space convolution}
\label{subsec:CS_convolution}

For the $2\to2$ CS subprocess, the hadronic differential cross
sections in $p_T$ and $y_{J/\psi}$ are obtained by integrating over
the rapidity of the recoiling gluon. Using the leading-order relation
$|\boldsymbol{k}_{T3}|=p_T$, the unpolarized differential cross section
can be written as
\begin{equation}
\frac{d\sigma_{\rm CS}}
{dp_T\,dy_{J/\psi}}
=
\int_{y_g^{\rm min}}^{y_g^{\rm max}}
dy_g\,
2p_T\,
\left[x_1g(x_1,\mu_F)\right]
\left[x_2g(x_2,\mu_F)\right]
\frac{d\hat{\sigma}_{\rm CS}}{d\hat{t}},
\label{eq:CS_sigma_pt_y}
\end{equation}
while the corresponding polarized cross section is
\begin{equation}
\frac{d\Delta\sigma_{\rm CS}}
{dp_T\,dy_{J/\psi}}
=
\int_{y_g^{\rm min}}^{y_g^{\rm max}}
dy_g\,
2p_T\,
\left[x_1\Delta g(x_1,\mu_F)\right]
\left[x_2\Delta g(x_2,\mu_F)\right]
\frac{d\Delta\hat{\sigma}_{\rm CS}}{d\hat{t}}.
\label{eq:CS_delta_sigma_pt_y}
\end{equation}
Here, $x_1$ and $x_2$ are determined for each phase-space point by
Eq.~\eqref{eq:x1x2_CS}, while the integration limits are fixed by the limits of longitudinal momentum fractions. The longitudinal double-spin asymmetry as a function of $p_T$ or
$y_{J/\psi}$ is then obtained from the corresponding differential
cross sections,
\begin{equation}
A_{LL}^{\rm CS}(p_T,y_{J/\psi})
=
\frac{
d\Delta\sigma_{\rm CS}/(dp_T\,dy_{J/\psi})
}{
d\sigma_{\rm CS}/(dp_T\,dy_{J/\psi})
}.
\label{eq:ALL_CS_pt_y}
\end{equation}

\subsection{Scales, PDFs, and theoretical uncertainties}
\label{subsec:scales_PDFs}

The factorization and renormalization scales are chosen to be equal
and are taken as
\begin{equation}
\mu_F=\mu_R
=
\sqrt{M^2+p_T^2}
=
\sqrt{4m_c^2+p_T^2},
\label{eq:scale_CS}
\end{equation}
and the perturbative scale dependence is estimated by varying the common
scale around its central value.

The numerical calculation is performed using the LHAPDF interface
with polarized and unpolarized gluon distributions. The nominal
prediction employs PDF sets at an order consistent with the
leading-order hard-scattering calculation. In addition, the LO
SDC is convoluted with NLO PDF sets to assess
the sensitivity of the predicted asymmetry to the perturbative order
of the PDFs. This comparison is intended only as a PDF-order
sensitivity test and should not be interpreted as a complete NLO
prediction, since the hard-scattering coefficient remains at LO.

The resulting theoretical uncertainty bands account for the
corresponding PDF uncertainties and variations of the factorization
and renormalization scales. Since the CS LDME cancels in the
single-channel ratio, its uncertainty does not contribute to the
prediction for $A_{LL}^{\rm CS}$.

\subsection{Numerical implementation}
\label{subsec:numerical_implementation}

The above formalism is implemented numerically to calculate the
longitudinal double-spin asymmetry for inclusive $J/\psi$ production
in longitudinally polarized proton-proton collisions at
$\sqrt{s}=27~\mathrm{GeV}$. For each $(p_T,y_{J/\psi})$ bin, the
recoil-gluon rapidity is integrated over the kinematically allowed
region, with $x_1$ and $x_2$ evaluated according to
Eq.~\eqref{eq:x1x2_CS}. The corresponding polarized and unpolarized
hard-scattering coefficients are evaluated using
Eqs.~\eqref{eq:CS_polarized} and \eqref{eq:CS_unpolarized},
respectively.

The numerical calculation provides the $p_T$- and
$y_{J/\psi}$-dependent CS contribution to $A_{LL}$, together with the
partonic analyzing power $\hat a_{LL}^{\rm CS}$, PDF uncertainties,
and perturbative scale variations. Since the calculation contains a
single color-singlet NRQCD channel, the resulting asymmetry is
independent of the overall normalization of the CS LDME. The CS
prediction therefore provides a theoretically clean benchmark for
assessing the sensitivity of inclusive $J/\psi$ production to the
polarized gluon distribution in the kinematic regime accessible at
SPD/NICA energies

\section{Results \& discussion}
\subsection{Channel-wise decomposition}

\begin{figure*}
	\centering
	\begin{minipage}[c]{0.98\textwidth}
		\includegraphics[width=7.5cm]{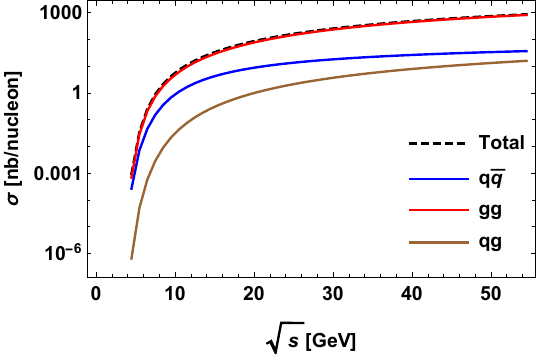}
	\end{minipage}
	\caption{\label{contrilog} Total $J/\psi$ cross section [nb/nucleon] and its channel-wise decomposition as a function of $\sqrt{s}$ in pN interactions \cite{Hsieh:2021yzg}.
}
\end{figure*}

Although this work focuses on the gluon-gluon initiated subprocess, it is important to assess the relative contributions of the different partonic channels to \(J/\psi\) production in the kinematic regime relevant for the SPD experiment. The contributions from the \(gg\), \(qg\), and \(q\bar q\) channels to \(J/\psi\) production within the NRQCD framework have previously been investigated as a function of the center-of-mass energy, with the \(gg\) channel found to provide the dominant contribution in the SPD/NICA energy range, as shown in Fig.~\ref{contrilog}~\cite{Hsieh:2021yzg}. In particular, our assessment of the relevant partonic channels in the kinematic region considered here confirms that the gluon-gluon initiated subprocess provides the dominant contribution at SPD energies. This dominance provides the basis for restricting the present analysis to the \(gg\) channel and, consequently, allows us to isolate the sensitivity of the longitudinal double-spin asymmetry to the gluon helicity distribution. Contributions from the \(qg\) and \(q\bar q\) channels, particularly in kinematic regions characterized by large momentum fractions or highly asymmetric partonic configurations, may nevertheless become relevant and warrant a dedicated quantitative study. A systematic treatment including all partonic subprocesses will therefore be considered in future work.

%
\begin{figure*}
	\centering
	\begin{minipage}[c]{0.98\textwidth}
		(a)\includegraphics[width=7.5cm]{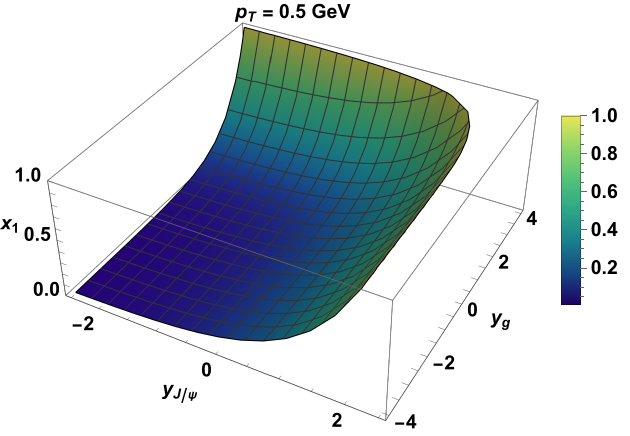}
		\hspace{0.03cm}
		(b)\includegraphics[width=7.5cm]{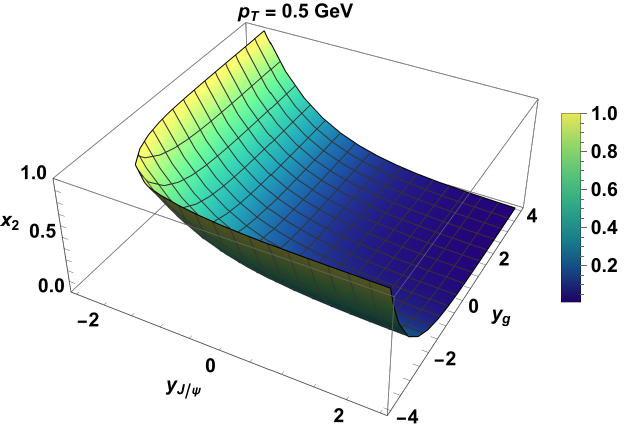}
		\hspace{0.03cm} \\
		(c)\includegraphics[width=7.5cm]{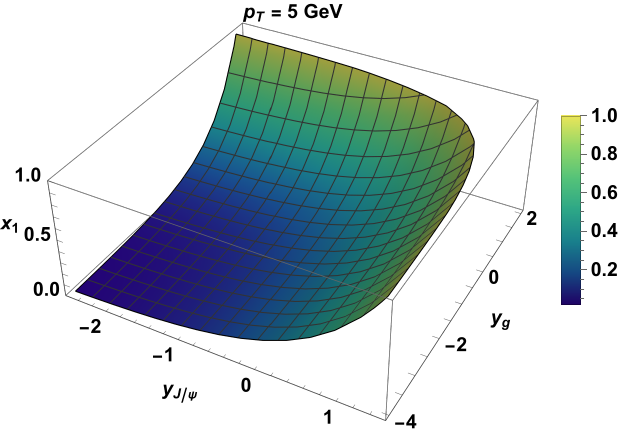}
		\hspace{0.03cm}
		(d)\includegraphics[width=7.5cm]{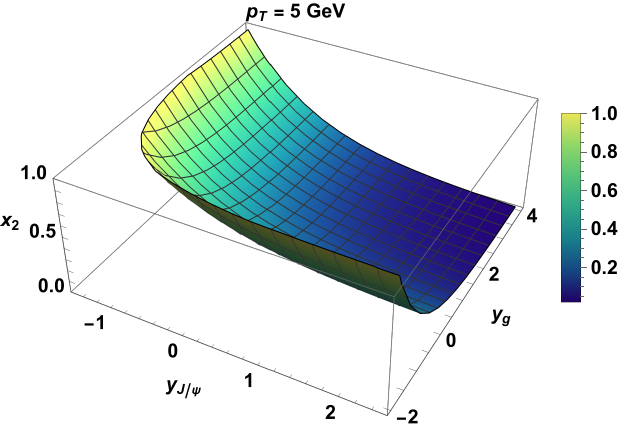}
		\hspace{0.03cm}
	\end{minipage}
	\caption{\label{fig_2to2_1} 
Longitudinal momentum fractions of the incoming partons for
the collinear $2\to2$ kinematics of $J/\psi$ production,
$g g\to J/\psi+g$, in $p p$ collisions at NICA energies. Panels
(a,c) show $x_1$, while panels (b,d) show $x_2$, as functions of the
$J/\psi$ rapidity $y_{J/\psi}$ and the recoil-gluon rapidity $y_g$ at
two fixed values of the $J/\psi$ transverse momentum $p_T$. The
momentum fractions are calculated from Eq.~\eqref{eq:x1x2_CS}, with
the physical phase space restricted by $x_1,x_2\leq1$.
	}
\end{figure*} 
\begin{figure*}
	\centering
	\begin{minipage}[c]{0.98\textwidth}
		(a)\includegraphics[width=7.5cm]{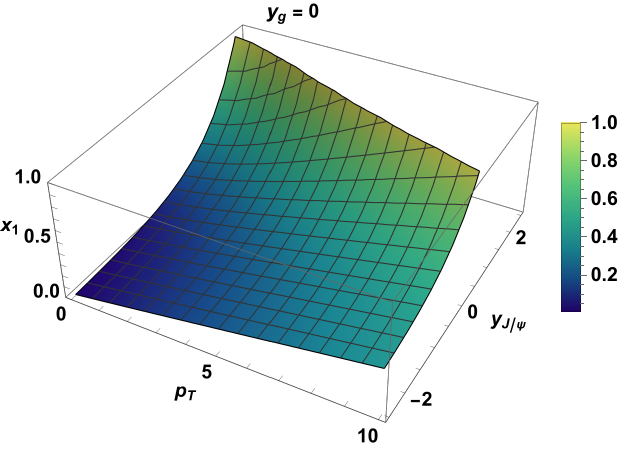}
		\hspace{0.03cm}
		(b)\includegraphics[width=7.5cm]{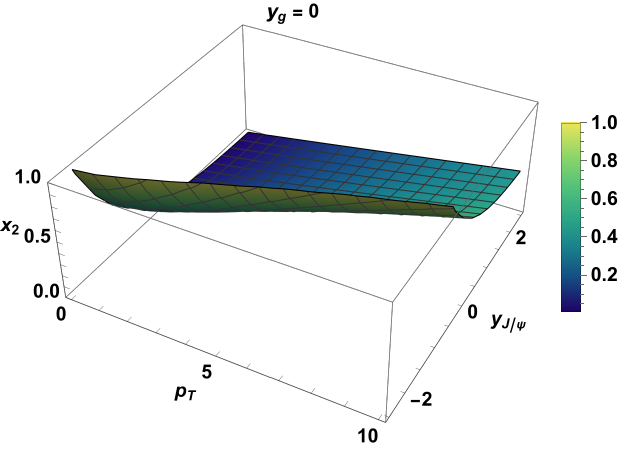}
		\hspace{0.03cm} \\
		(c)\includegraphics[width=7.5cm]{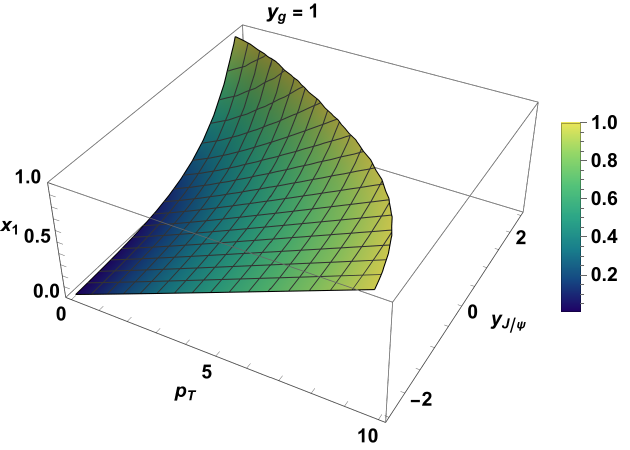}
		\hspace{0.03cm}
		(d)\includegraphics[width=7.5cm]{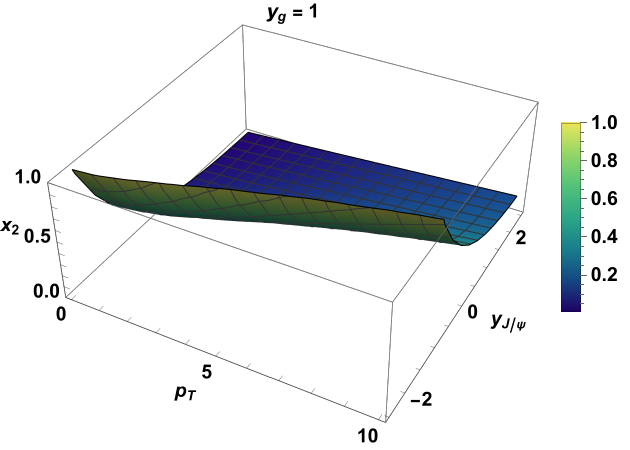}
		\hspace{0.03cm}
	\end{minipage}
	\caption{\label{fig_2to2_2} 
Longitudinal momentum fractions of the incoming partons for
the collinear $2\to2$ kinematics of $J/\psi$ production,
$g g\to J/\psi+g$, in $p p$ collisions at NICA energies. Panels
(a,c) show $x_1$, while panels (b,d) show $x_2$, as functions of the
$J/\psi$ rapidity $y_{J/\psi}$ and transverse momentum $p_T$ at two
fixed values of the recoil-gluon rapidity $y_g$. The momentum
fractions are obtained from Eq.~\eqref{eq:x1x2_CS}, with the allowed
phase space determined by $x_1,x_2\leq1$.
	}
\end{figure*} 
\begin{figure*}
	\centering
	\begin{minipage}[c]{0.98\textwidth}
		(a)\includegraphics[width=7.5cm]{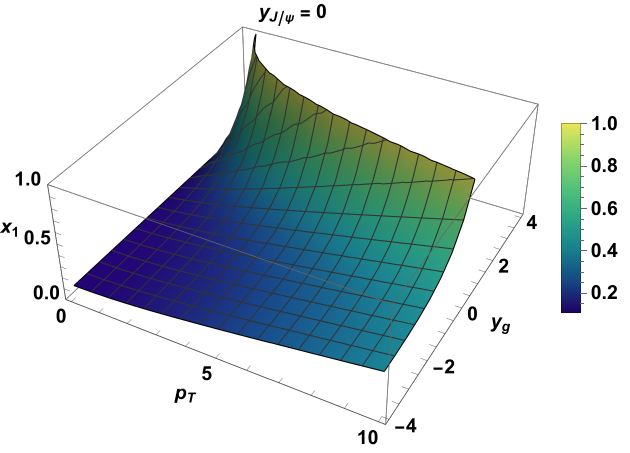}
		\hspace{0.03cm}
		(b)\includegraphics[width=7.5cm]{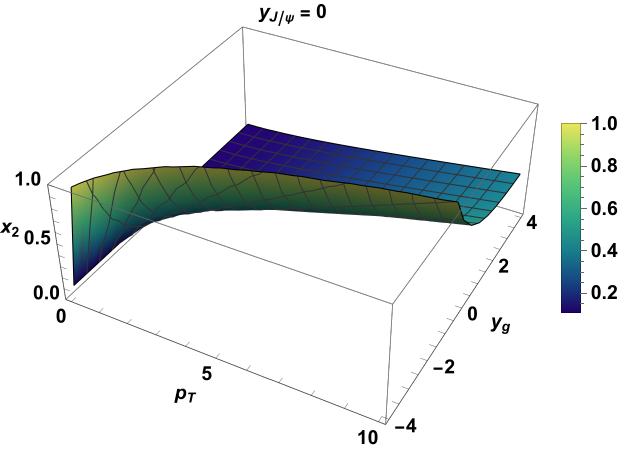}
		\hspace{0.03cm} \\
		(c)\includegraphics[width=7.5cm]{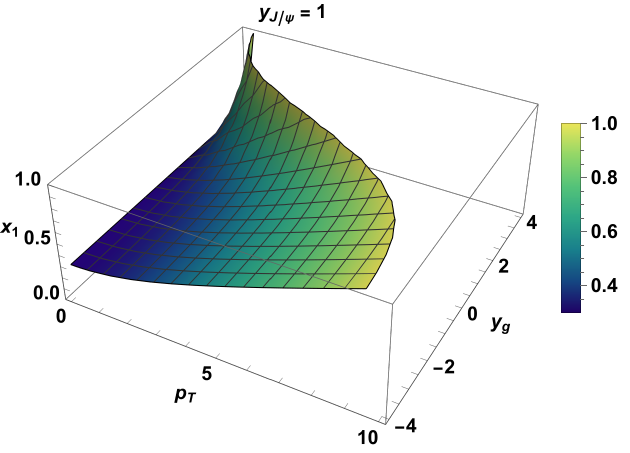}
		\hspace{0.03cm}
		(d)\includegraphics[width=7.5cm]{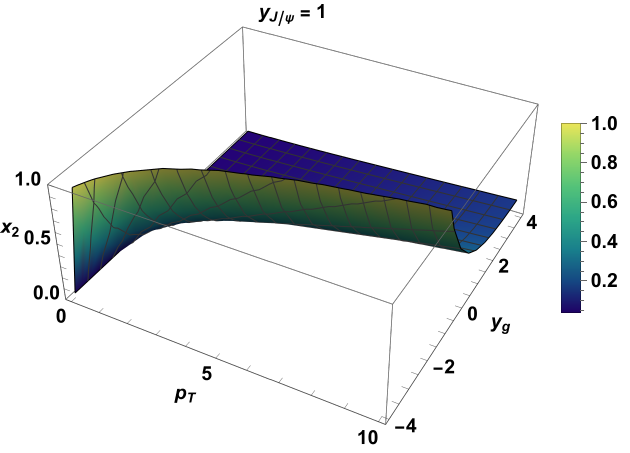}
		\hspace{0.03cm}
	\end{minipage}
	\caption{\label{fig_2to2_3} 
Longitudinal momentum fractions of the incoming partons for
the collinear $2\to2$ kinematics of $J/\psi$ production,
$g g\to J/\psi+g$, in $p p$ collisions at NICA energies. Panels
(a,c) show $x_1$, while panels (b,d) show $x_2$, as functions of the
recoil-gluon rapidity $y_g$ and the $J/\psi$ transverse momentum $p_T$
at two fixed values of the $J/\psi$ rapidity $y_{J/\psi}$. The
momentum fractions are calculated from Eq.~\eqref{eq:x1x2_CS}, with
the physical phase space restricted by $x_1,x_2\leq1$.
	}
\end{figure*} 
\subsection{Kinematic distributions}
%
%
%
%
To study the $2\to2$ kinematics, we consider three complementary
choices of independent variables, shown in
Figs.~\ref{fig_2to2_1}-\ref{fig_2to2_3}. In all cases, the momentum
fractions of the incoming partons are obtained from the kinematic
relations in Eq.~\eqref{eq:x1x2_CS}, with the physical phase space
restricted by $0<x_1,x_2\leq1$.


Figure~\ref{fig_2to2_1} shows the longitudinal momentum fractions of the incoming partons, (a,c) $x_1$ and (b,d) $x_2$, as functions of the $J/\psi$ rapidity $y_{J/\psi}$ and the recoil-gluon rapidity $y_g$ at fixed values of the $J/\psi$ transverse momentum $p_T$. The distributions exhibit the expected reflection symmetry under 
\begin{equation}
(y_{J/\psi},y_g)\rightarrow
(-y_{J/\psi},-y_g),
\qquad
x_1\leftrightarrow x_2,
\end{equation}
which follows directly from the corresponding expressions in Eq.~\eqref{eq:x1x2_CS}.

At fixed $p_T$ and $y_g$, increasing $y_{J/\psi}$ enhances $x_1$ and
suppresses $x_2$, whereas decreasing $y_{J/\psi}$ has the opposite
effect. Similarly, at fixed $p_T$ and $y_{J/\psi}$, increasing $y_g$
enhances $x_1$ and suppresses $x_2$, while decreasing $y_g$ leads to
the opposite behavior. Thus, the parton moving in the direction of
the corresponding final-state particle carries the larger longitudinal
momentum fraction.

When the fixed value of $p_T$ is increased, as illustrated by the
comparison of panels (a,b) with panels (c,d), the values of both
$x_1$ and $x_2$ increase through their dependence on the transverse
mass $m_T=\sqrt{M^2+p_T^2}$. Consequently, the region of
$(y_{J/\psi},y_g)$ phase space satisfying $x_1,x_2\leq1$ becomes more
restricted at larger $p_T$. At central rapidities,
$y_{J/\psi}=y_g=0$, the two momentum fractions are equal,
\begin{equation}
x_1=x_2=\frac{m_T+p_T}{\sqrt{s}},
\end{equation}
as follows directly from Eq.~\eqref{eq:x1x2_CS}.



Figure~\ref{fig_2to2_2} shows the longitudinal momentum fractions
$x_1$ and $x_2$, displayed in panels (a,c) and (b,d), respectively, as
functions of the $J/\psi$ rapidity $y_{J/\psi}$ and transverse momentum
$p_T$, for fixed values of the recoil-gluon rapidity $y_g$. The
reflection symmetry
\begin{equation}
y_{J/\psi}\rightarrow -y_{J/\psi},
\qquad
x_1\leftrightarrow x_2,
\end{equation}
is again evident and follows directly from
Eq.~\eqref{eq:x1x2_CS}, together with the interchange of the two beam
directions.

At fixed $p_T$ and $y_g$, increasing $y_{J/\psi}$ increases $x_1$ and
decreases $x_2$. At fixed $y_g$ and $y_{J/\psi}$, both $x_1$ and $x_2$ increase
with increasing $p_T$, owing to the increase of the $J/\psi$ transverse
mass. The allowed region in the $(p_T,y_{J/\psi})$ plane is therefore
progressively restricted as $p_T$ increases, particularly at larger
recoil gluon rapidities, where one of the momentum fractions approaches
the kinematic limit $x_i=1$ quickly. At $y_{J/\psi}=y_g=0$, the two momentum
fractions become equal,
\begin{equation}
x_1=x_2=\frac{m_T+p_T}{\sqrt{s}}.
\end{equation}
%

Figure~\ref{fig_2to2_3} presents the longitudinal momentum fractions $x_1$ and $x_2$, shown in panels (a,c) and (b,d), respectively, as functions of the recoil gluon rapidity $y_g$ and the $J/\psi$ transverse momentum $p_T$, for fixed values of the $J/\psi$ rapidity $y_{J/\psi}$. The distributions exhibit the corresponding reflection symmetry under
\begin{equation}
y_g\rightarrow -y_g,
\qquad
x_1\leftrightarrow x_2,
\end{equation}
when the beam directions are interchanged, as dictated by Eq.~\eqref{eq:x1x2_CS}.

At fixed $y_{J/\psi}$ and $y_g$, both $x_1$ and $x_2$ increase with
increasing $p_T$, reflecting the increase of the transverse mass of
the $J/\psi$. As the magnitude of the fixed $y_{J/\psi}$ is increased,
the allowed range of $p_T$ becomes progressively narrower because the
condition $x_1,x_2\leq1$ is reached at smaller transverse momentum.
This reduction of the available $p_T$ range is a direct consequence of
the finite center-of-mass energy and becomes particularly relevant for
the relatively low collision energies considered here.

At fixed $p_T$ and $y_{J/\psi}$, increasing $y_g$ increases $x_1$ and
decreases $x_2$. Consequently, the recoil-gluon rapidity provides an
additional degree of freedom in the $2\to2$ kinematics and directly
controls how the longitudinal momentum required from the two incoming
partons is shared between the two proton beams. The numerical values of $x_1, x_2$ complimentary to discussion for Figs.~\ref{fig_2to2_1}-\ref{fig_2to2_3} has been listed in Table~\ref{table2to2} in the Appendix.






\subsection{Longitudinal double-spin asymmetry}

For the numerical evaluation, we use the \texttt{LHAPDF} framework~\cite{Buckley:2014ana} together with several polarized and unpolarized parton distribution function (PDF) sets. For the polarized gluon distributions, we employ the NNPDFpol2.0 sets at LO and NLO, namely \texttt{NNPDFpol20\_lo\_as\_01180} and \texttt{NNPDFpol20\_nlo\_as\_01180}, respectively~\cite{Cruz-Martinez:2025ahf}. For the corresponding unpolarized distributions, we use the NNPDF3.1 LO and NLO sets, \texttt{NNPDF31\_lo\_as\_0118} and \texttt{NNPDF31\_nlo\_as\_0118}~\cite{NNPDF:2017mvq}. As an independent comparison, we also employ the CT18 family, using \texttt{CT18LO}~\cite{Yan:2022pzl} and \texttt{CT18NLO}~\cite{Hou:2019qau} for the unpolarized LO and NLO calculations, respectively. We choose these combinations to separate the different $A_{LL}$ curves on the basis of LO and NLO. We understand that it is ideal to take the two PDFs of the same type while calculating the asymmetry but still we consider CT18LO and CT18NLO as we were facing some discrepancies while working with the replica with transverse momentum values of NNPDF31.The hard-scattering coefficient is evaluated at leading order, while the different PDF orders are used to assess the sensitivity of the asymmetry to the choice of parton distributions~\cite{Klasen:2003zn}.
%
%
%
%
\begin{figure*}
	\centering
	\begin{minipage}[c]{0.98\textwidth}
		(a)\includegraphics[width=7.5cm]{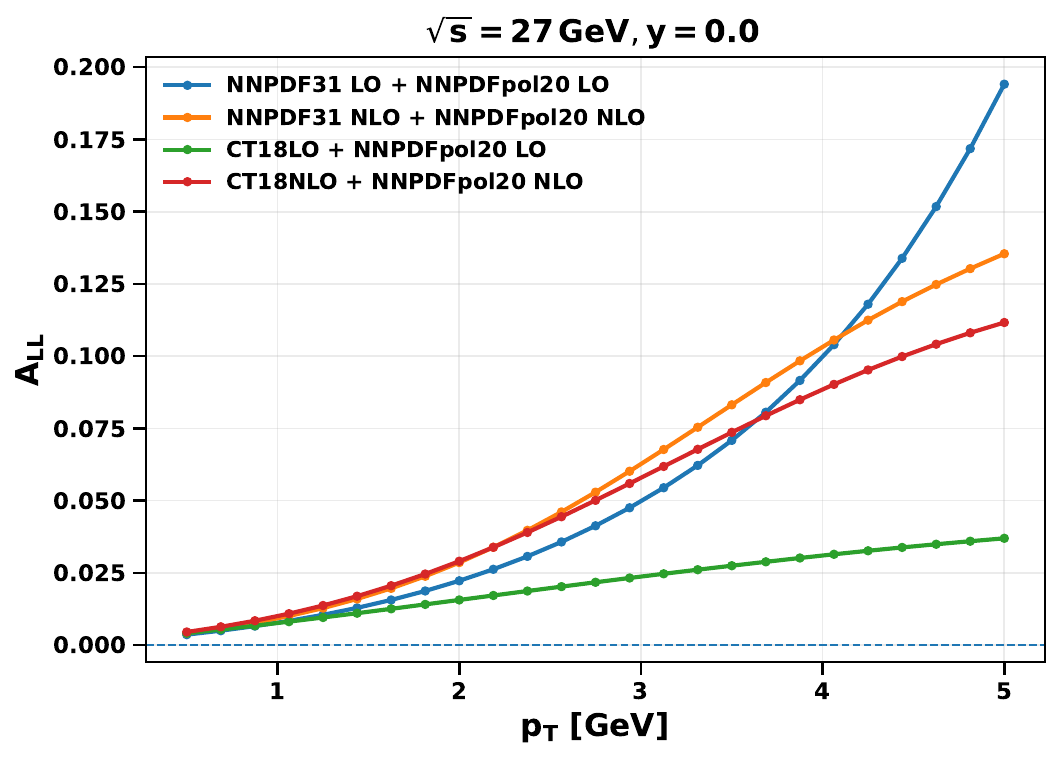}
		\hspace{0.03cm}
		(b)\includegraphics[width=7.5cm]{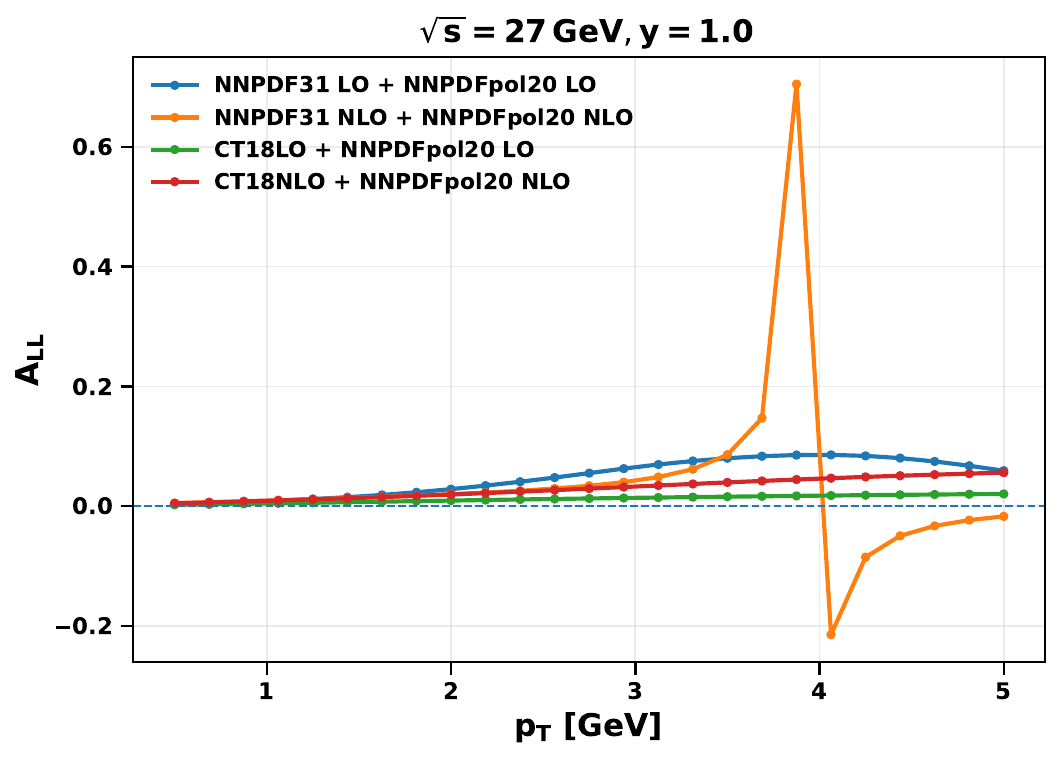}
		\hspace{0.03cm} \\
	\end{minipage}
	\caption{\label{fig:ALL_pt_pdf}  
Longitudinal double-spin asymmetry \(A_{LL}\) for color-singlet \(J/\psi\) production through \(g g\to J/\psi+g\) as a function of the transverse momentum \(p_T\) at central rapidity, (a) \(y_{J/\psi}=0\) and (b) \(y_{J/\psi}=1\), for \(\sqrt{s}=27\) GeV. The predictions obtained with four combinations of unpolarized and polarized PDFs are shown to assess the dependence of the spin asymmetry on the PDF set and perturbative PDF order. The hard-scattering coefficient is evaluated at leading order. 
}
\end{figure*} 
%
\subsection{Dependence on the transverse momentum and PDF input}
\label{subsec:pt_pdf}

We first examine the transverse-momentum dependence of $A_{LL}$ at
central and forward rapidities.  Figure~\ref{fig:ALL_pt_pdf} shows the
results at $y_{J/\psi}=0$ and $y_{J/\psi}=1$ for the four combinations
of polarized and unpolarized PDFs considered in this work.  A
pronounced dependence on the PDF input is already visible at moderate
and large transverse momentum, demonstrating that the asymmetry is not
controlled solely by the hard-scattering coefficient.

At central rapidity, the four predictions exhibit a positive and
monotonically increasing asymmetry over the displayed transverse-momentum
range.  At low $p_T$, the different PDF combinations give rather similar
results, with $A_{LL}$ of only a magnitude of $10^{-3}$. The separation
between the predictions becomes progressively larger as $p_T$ increases.
For the NNPDF31 LO-NNPDFpol20 LO combination, the asymmetry rises to
approximately $0.19$ at $p_T=5$~GeV, whereas the corresponding
NNPDF31 NLO-NNPDFpol20 NLO result reaches about $0.14$.  The CT18LO
combination gives a substantially smaller asymmetry, remaining below
$0.04$ over the displayed range, while the CT18NLO result reaches about
$0.11$.  Thus, at central rapidity, the difference between the PDF
families becomes particularly significant in the high-$p_T$ region.

This behavior can be understood qualitatively from the kinematics of
the process.  Increasing $p_T$ at fixed rapidity requires larger
longitudinal momentum fractions from the incoming partons.  At
$\sqrt{s}=27$~GeV this rapidly moves the process towards the
moderate-to-large-$x$ gluon region, where the polarized gluon
distribution is less constrained and where different global PDF
analyses can give substantially different predictions. Consequently,
the sensitivity of $A_{LL}$ to the PDF choice increases with $p_T$.
The result is particularly relevant for a future SPD measurement:
rather than merely providing a test of the hard-scattering description,
the transverse-momentum dependence of the asymmetry can discriminate
between different assumptions about the polarized gluon density.

At $y_{J/\psi}=1$, the qualitative behavior at low and moderate $p_T$
is similar, but the PDF dependence becomes more pronounced.  The
NNPDF31 LO-NNPDFpol20 LO prediction increases from values close to zero
at low $p_T$ to approximately $0.08$ around $p_T\simeq4$~GeV and then
decreases mildly.  The CT18LO prediction remains much smaller, at the
level of a few percent.  The CT18NLO result follows a comparatively
smooth positive behavior and reaches roughly $0.05$ at the upper end
of the displayed range.

The NNPDF31 NLO-NNPDFpol20 NLO result, however, develops a very large
excursion around $p_T\simeq4$~GeV, including a change of sign.  The
magnitude of this feature is clearly inconsistent with the otherwise
smooth behavior of the other PDF combinations and with the neighboring
kinematic points.  It should therefore not be interpreted as a genuine
enhancement of the physical asymmetry.  Such behavior can arise when
the unpolarized denominator becomes anomalously small or changes sign
in individual PDF replicas in a kinematic region where the
corresponding PDF set is poorly constrained.  We consequently regard
this narrow structure as a numerical/PDF-instability region rather
than as a robust prediction.  A conservative phenomenological
interpretation of the NLO-PDF result should therefore be restricted to
the region in which the denominator remains stable and the replica
predictions give a well-defined asymmetry.

The comparison in Fig.~\ref{fig:ALL_pt_pdf} also shows an important
distinction between PDF order and hard-scattering order.  Since the
hard coefficient is kept at leading order throughout this analysis,
the curves labelled ``NLO'' do not represent a complete NLO prediction
for $A_{LL}$.  They instead quantify the sensitivity of the LO
observable to replacing LO PDFs by NLO PDF sets.  The sizeable
difference between the LO- and NLO-PDF results at moderate and large
$p_T$ consequently provides an estimate of the perturbative-PDF
scheme dependence, rather than an NLO correction to the partonic
cross section itself.
%
%
%
\begin{figure*}
	\centering
	\begin{minipage}[c]{0.98\textwidth}
		(a)\includegraphics[width=7.5cm]{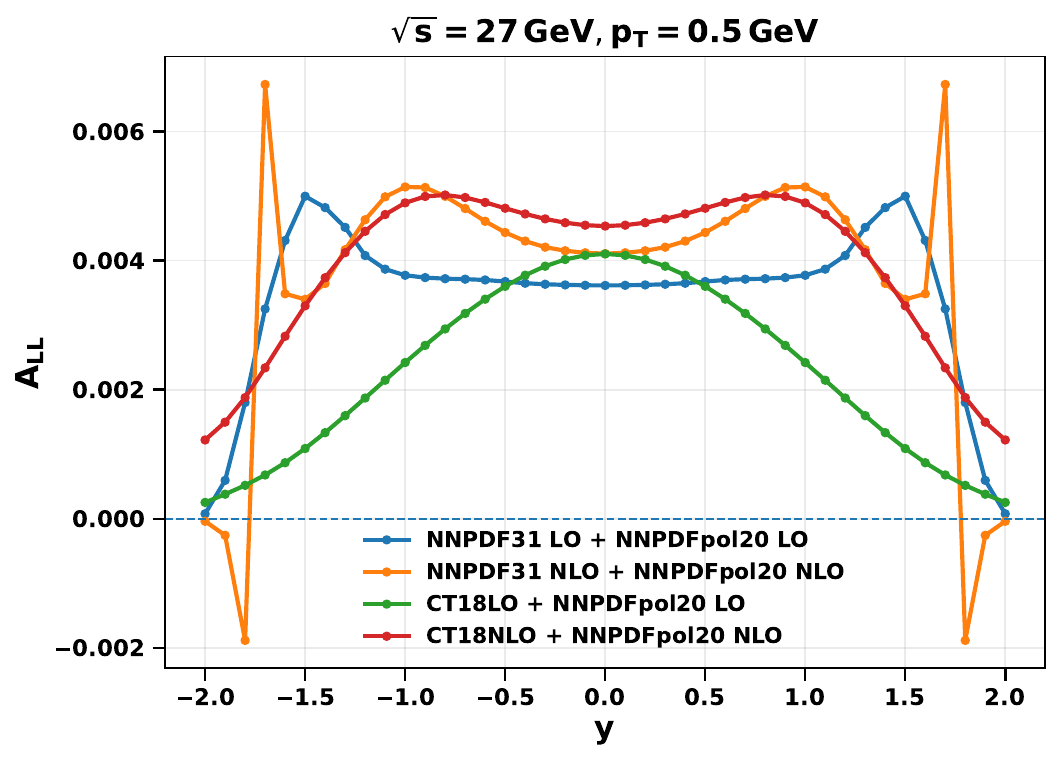}
		\hspace{0.03cm}
		(b)\includegraphics[width=7.5cm]{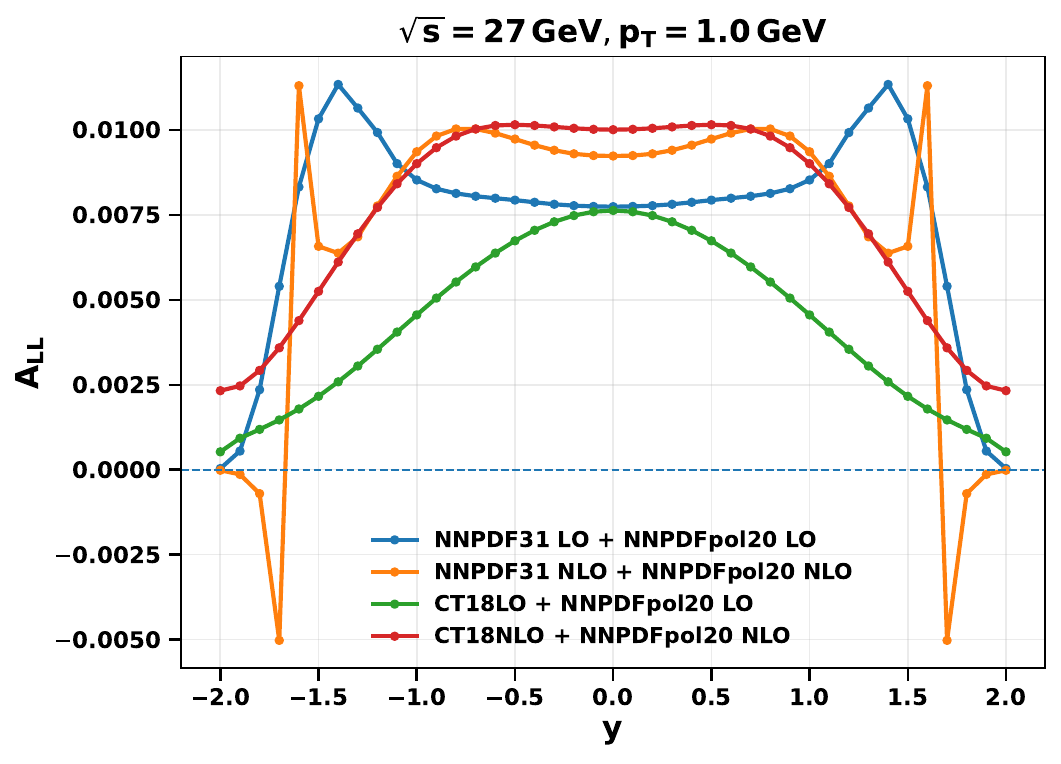}
		\hspace{0.03cm} 
        (c)\includegraphics[width=7.5cm]{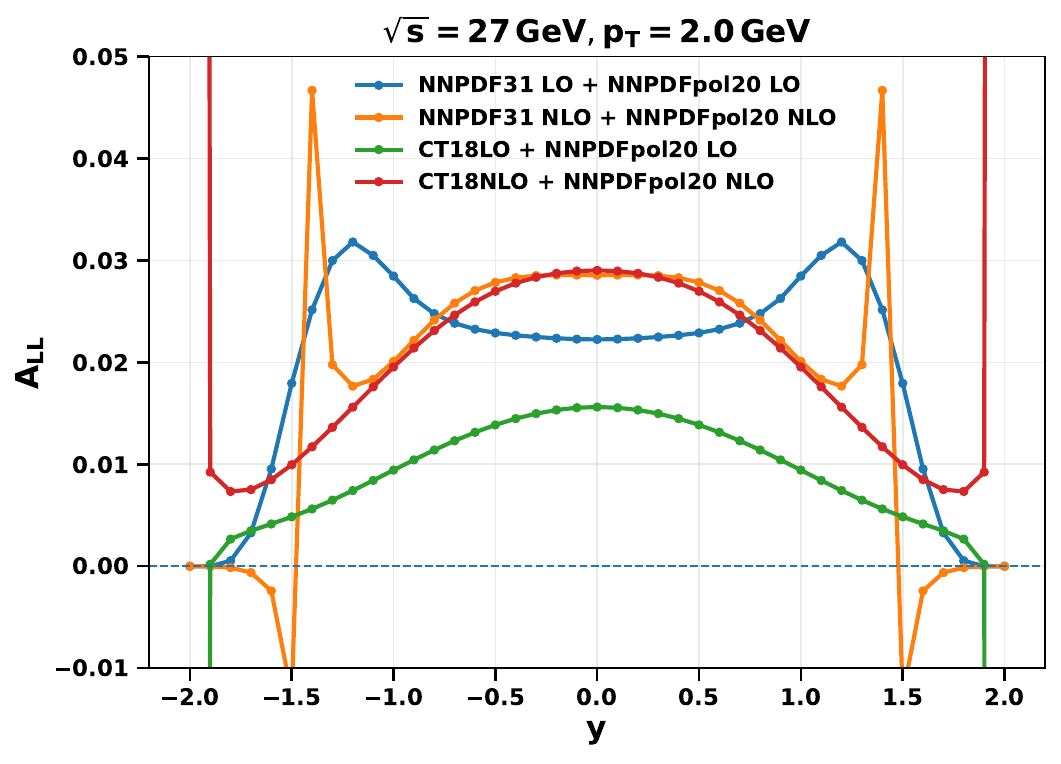}
		\hspace{0.03cm} \\
	\end{minipage}
	\caption{\label{fig:ALL_y_pdf}  
Longitudinal double-spin asymmetry \(A_{LL}\) for color-singlet \(J/\psi\) production through \(g g\to J/\psi+g\) at \(\sqrt{s}=27\) GeV as a function of \(y_{J/\psi}\) at (a) \(p_T=0.5\) GeV, (b) \(p_T=1.0\) GeV, and (c) \(p_T=2.0\) GeV. The results are shown using the NNPDF31 LO and NNPDFpol20 LO, NNPDF31 NLO and NNPDFpol20 NLO, CT18LO and NNPDFpol20 LO, and CT18NLO and NNPDFpol20 NLO PDF combinations~\cite{Cruz-Martinez:2025ahf,NNPDF:2017mvq,Yan:2022pzl,Hou:2019qau}. The hard-scattering coefficient is evaluated at leading order, while the different PDF orders are used to assess the sensitivity of the asymmetry to the choice of parton distributions.
}
\end{figure*} 

\subsection{Rapidity dependence}
\label{subsec:y_pdf}

The complementary rapidity dependence is shown in
Fig.~\ref{fig:ALL_y_pdf} for $p_T=0.5$, $1.0$, and $2.0$~GeV.  The
rapidity dependence provides a particularly useful view of the
kinematic symmetry of the polarized $p p$ collision.  Because the two
incoming protons are identical, the observable is expected to be
symmetric under the interchange of the two beam directions,
corresponding to
\begin{equation}
 A_{LL}(y_{J/\psi})=A_{LL}(-y_{J/\psi}) .
\end{equation}
The calculated results exhibit this symmetry to a very good accuracy
away from the kinematic boundaries.

At $p_T=0.5$~GeV, the central predictions remain positive over the
central rapidity region and are relatively flat around midrapidity,
with $A_{LL}$ of order $4\times10^{-3}$. For CT18NLO and NNPDFpol20 NLO sets the asymmetry at intermediate rapidities increases by $25\%$ than its value at mid-rapidities. For NNPDF31 LO and NNPDFpol20 LO, NNPDF31 NLO and NNPDFpol20 NLO sets, the asymmetry increases towards intermediate rapidities, reaching values of roughly $(5$-$6)\times10^{-3}$ before decreasing rapidly as
$|y_{J/\psi}|$ approaches the kinematic boundary. For CT18LO and NNPDFpol20 LO the asymmetry falls as the rapidity increases.

The overall magnitude is small at this transverse momentum, indicating that the
spin-dependent and unpolarized gluon convolutions remain rather
similar in the predominantly small-to-moderate-$x$ region probed near
midrapidity.

Increasing the transverse momentum enhances the magnitude of the
asymmetry.  At $p_T=1$~GeV, the central values around midrapidity are
approximately $(8$-$10)\times10^{-3}$, while the intermediate-rapidity
region reaches values of order $10^{-2}$.  At $p_T=2$~GeV, the
midrapidity asymmetry rises further to approximately $2.5\times10^{-2}$,
with values approaching $3\times10^{-2}$ at intermediate rapidity.
Thus, the $p_T$ and rapidity dependences are correlated: increasing
$p_T$ moves the process towards larger partonic momentum fractions,
while moving away from midrapidity makes the momentum fractions of the
two incoming gluons increasingly asymmetric.  Both effects enhance the
sensitivity to the polarized gluon distributions.

The comparison among PDF combinations reveals that the qualitative
rapidity dependence is considerably more stable than the absolute
normalization at larger $p_T$.  For $p_T=0.5$~GeV, the four PDF
combinations remain close to one another over the central rapidity
interval.  At $p_T=1$~GeV, the NNPDF31 LO and NNPDF31 NLO combinations
give values around the $10^{-2}$ level in the central region, while
the CT18LO prediction is systematically lower.  The CT18NLO result
remains closer to the NNPDF predictions.  At $p_T=2$~GeV, the
separation becomes more pronounced: the CT18LO prediction is
approximately a factor of two smaller than the NNPDF31 LO prediction
around midrapidity, whereas the NNPDF31 NLO and CT18NLO results are
considerably closer to one another.

A characteristic feature of all four calculations is the rapid
suppression of the asymmetry as $|y_{J/\psi}|$ approaches the
kinematic limit.  This is accompanied in several cases by sharp
excursions very close to the endpoints.  The endpoint region
corresponds to highly asymmetric partonic momentum fractions and a
rapidly decreasing unpolarized cross section.  Ratios such as
$A_{LL}=\Delta\sigma/\sigma$ can consequently become numerically
unstable when the denominator becomes very small.  The endpoint
structures should therefore be treated with caution and should not be
used to infer a physical enhancement of the spin asymmetry.  The
central rapidity region, where the denominator is well behaved, gives
a substantially more robust prediction.

%
\begin{figure*}
	\centering
	\begin{minipage}[c]{0.98\textwidth}
		(a)\includegraphics[width=7.5cm]{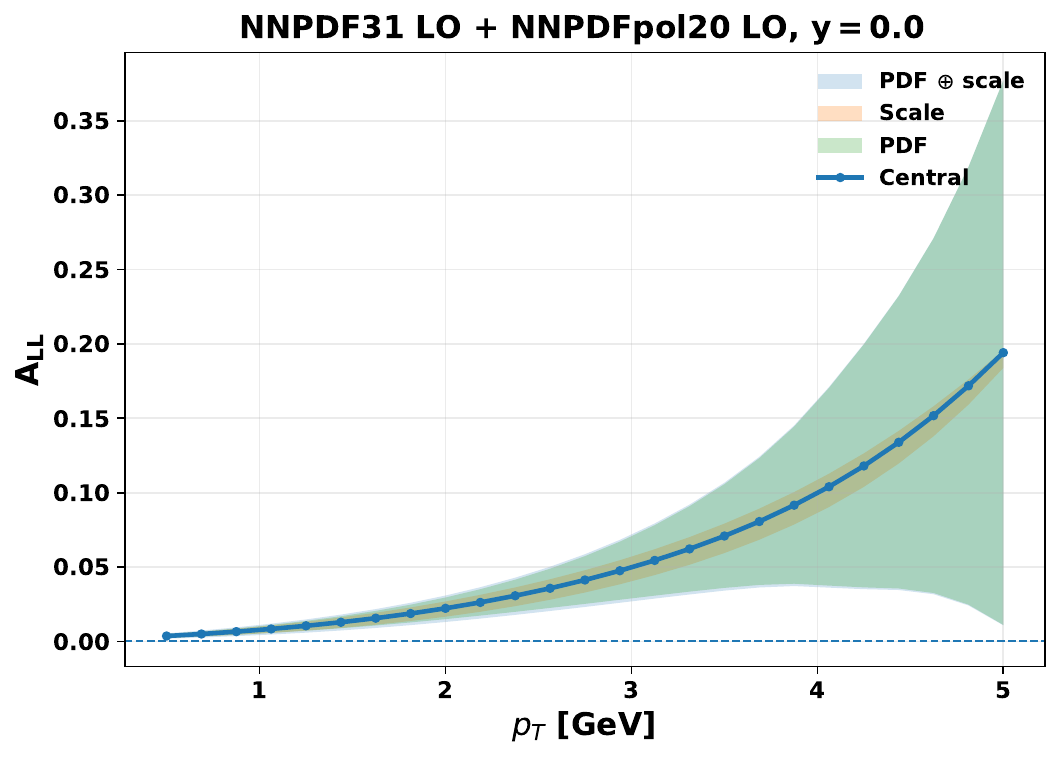}
		\hspace{0.03cm}
		(b)\includegraphics[width=7.5cm]{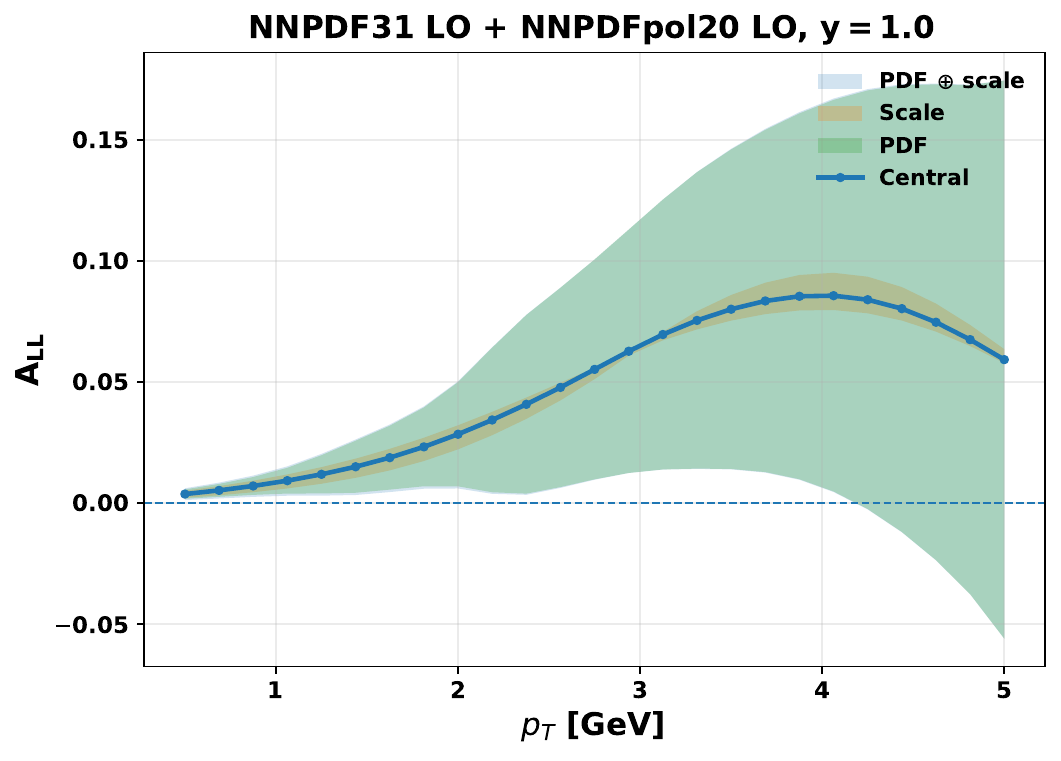}
		\hspace{0.03cm} \\
	\end{minipage}
	\caption{\label{fig:ALL_pt_unc}  
    Longitudinal double-spin asymmetry \(A_{LL}\) for color-singlet \(J/\psi\) production through \(g g\to J/\psi+g\) as a function of \(p_T\) at (a) \(y_{J/\psi}=0\), (b) \(y_{J/\psi}=1\) and \(\sqrt{s}=27\) GeV. The central prediction employs the NNPDF31 LO and NNPDFpol20 LO PDF sets. The separate PDF and scale uncertainty bands, as well as their combined uncertainty, are shown to quantify the theoretical uncertainty of the predicted spin asymmetry in the forward-rapidity region. 
}
\end{figure*} 
%
\begin{figure*}
	\centering
	\begin{minipage}[c]{0.98\textwidth}
		(a)\includegraphics[width=7.5cm]{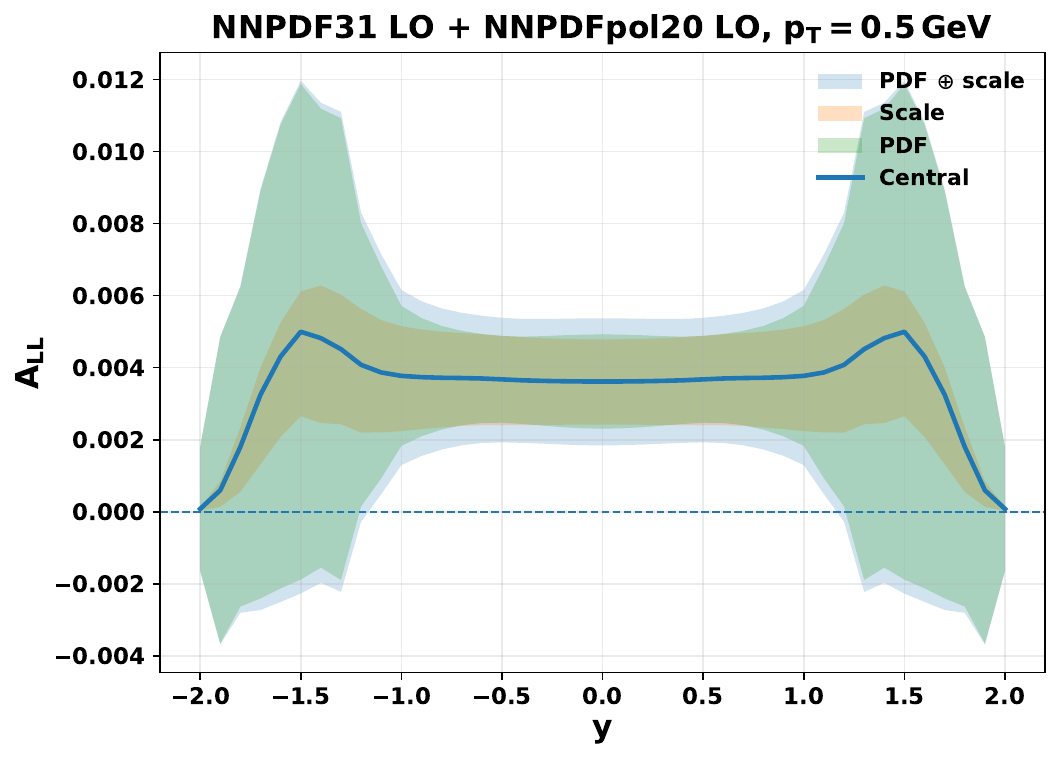}
		\hspace{0.03cm}
		(b)\includegraphics[width=7.5cm]{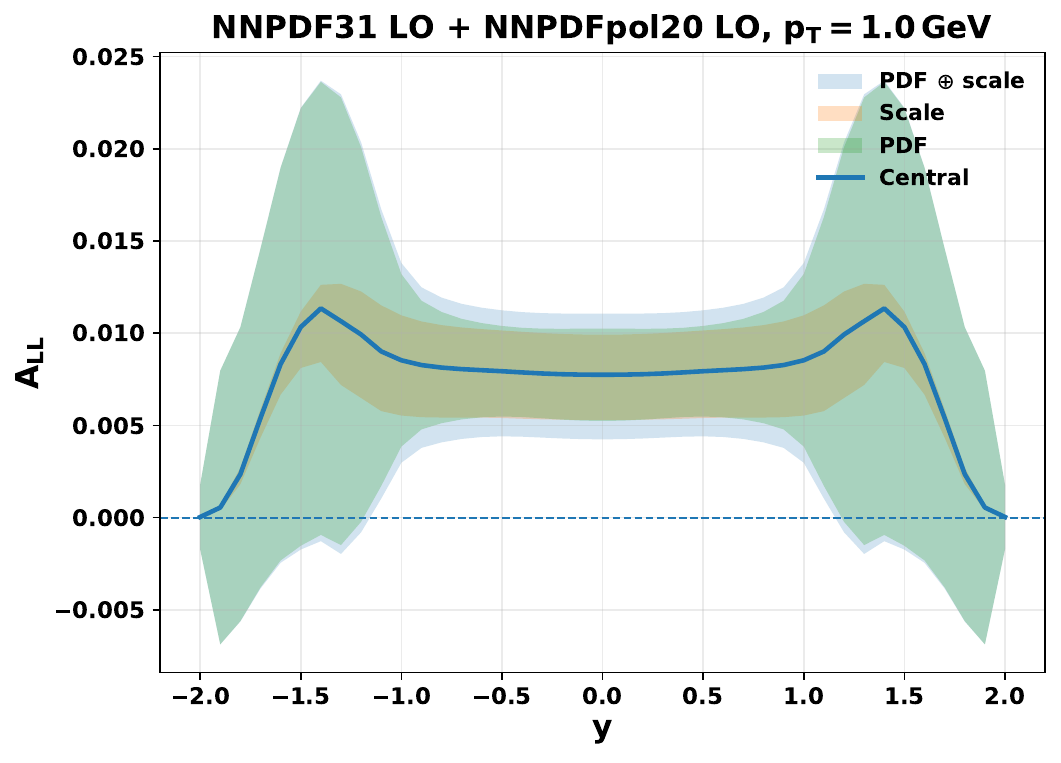}
		\hspace{0.03cm} 
        (c)\includegraphics[width=7.5cm]{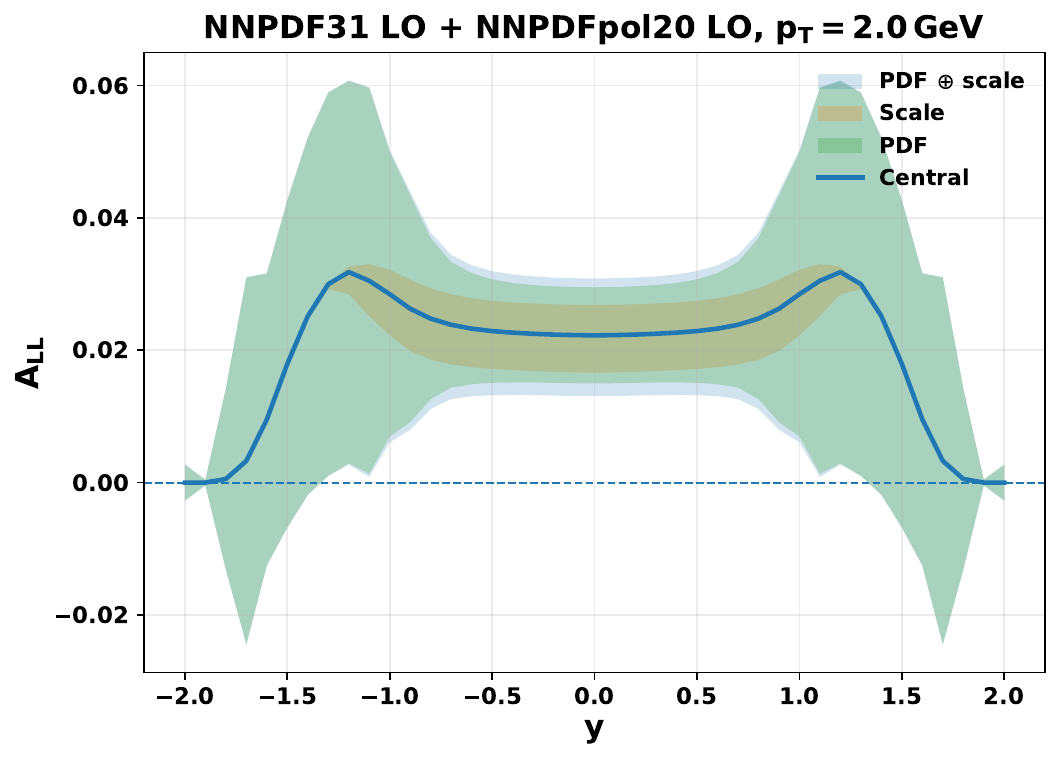}
		\hspace{0.03cm} \\
	\end{minipage}
	\caption{\label{fig:ALL_y_unc}  
Longitudinal double-spin asymmetry \(A_{LL}\) for color-singlet \(J/\psi\) production through \(g g\to J/\psi+g\) at \(\sqrt{s}=27\) GeV as a function of \(y_{J/\psi}\) at (a) \(p_T=0.5\) GeV, (b) \(p_T=1.0\) GeV, and (c) \(p_T=2.0\) GeV. The central prediction is obtained with the NNPDF31 LO and NNPDFpol20 LO PDF sets. The displayed bands represent the PDF uncertainty, the uncertainty from factorization- and renormalization-scale variations, and the total uncertainty obtained by combining the two sources in quadrature.
}
\end{figure*} 
%
\begin{figure*}
	\centering
	\begin{minipage}[c]{0.98\textwidth}
		(a)\includegraphics[width=7.5cm]{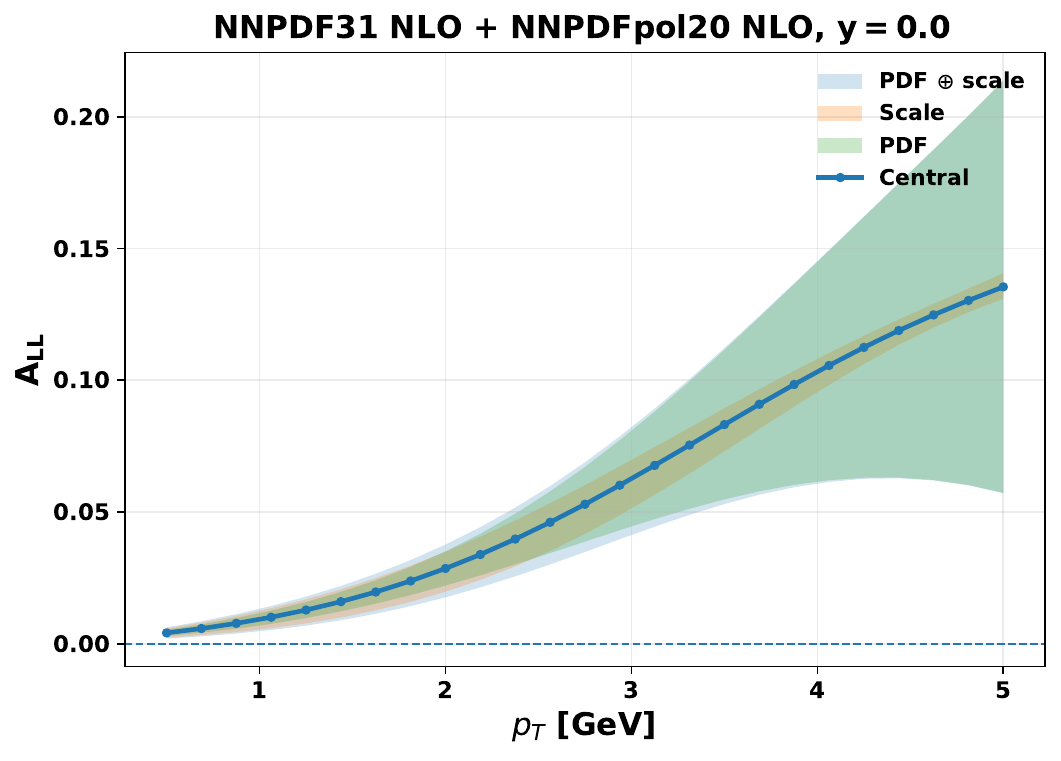}
		\hspace{0.03cm}
		(b)\includegraphics[width=7.5cm]{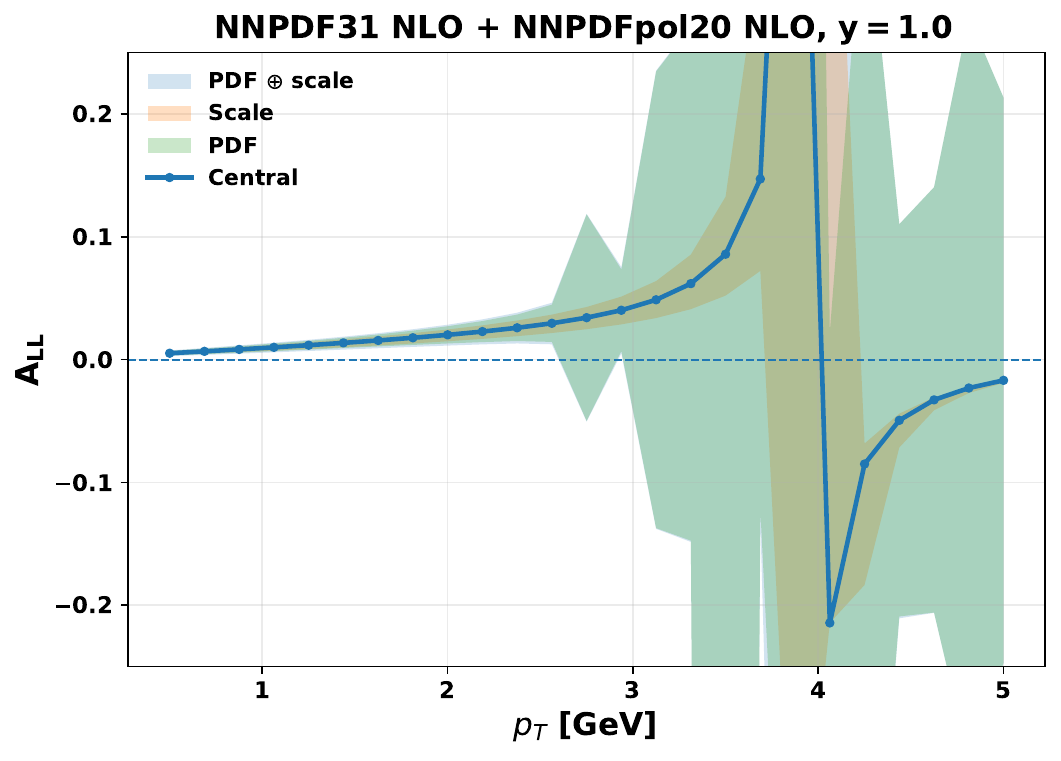}
		\hspace{0.03cm} \\
	\end{minipage}
	\caption{\label{fig:ALL_pt_uncNLO}  
    Longitudinal double-spin asymmetry \(A_{LL}\) for color-singlet \(J/\psi\) production through \(g g\to J/\psi+g\) as a function of \(p_T\) at (a) \(y_{J/\psi}=0\), (b) \(y_{J/\psi}=1\) and \(\sqrt{s}=27\) GeV. The central prediction employs the NNPDF31 NLO and NNPDFpol20 NLO PDF sets. The separate PDF and scale uncertainty bands, as well as their combined uncertainty, are shown to quantify the theoretical uncertainty of the predicted spin asymmetry in the forward-rapidity region. 
}
\end{figure*} 
%
\begin{figure*}
	\centering
	\begin{minipage}[c]{0.98\textwidth}
		(a)\includegraphics[width=7.5cm]{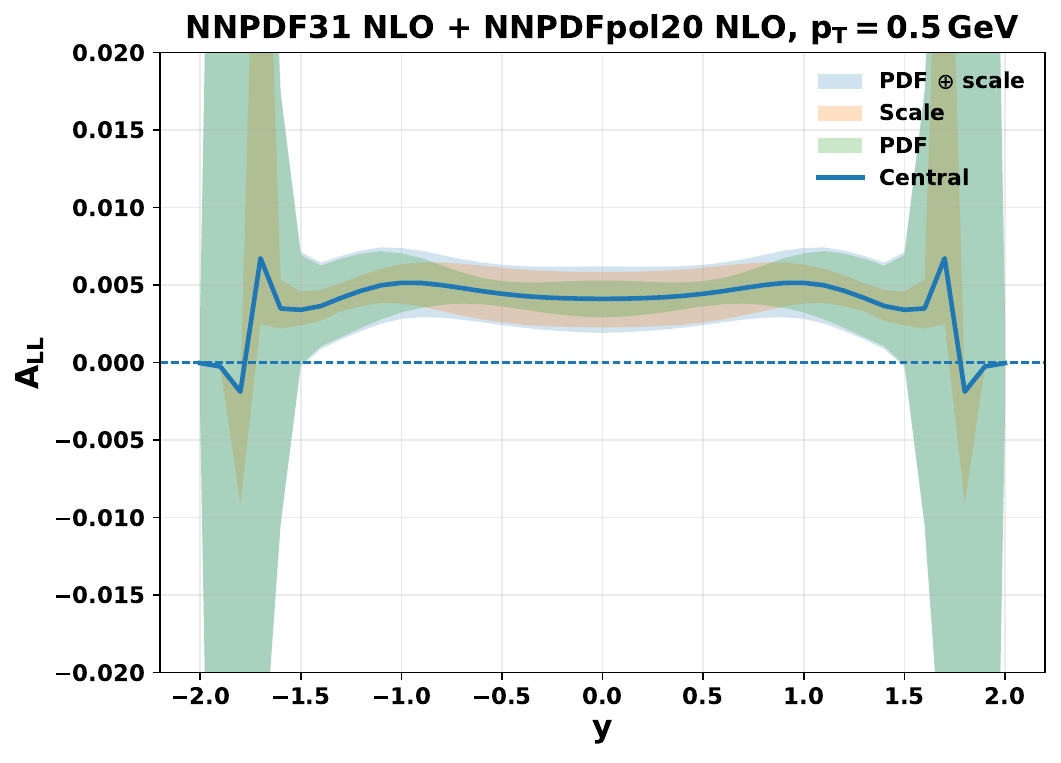}
		\hspace{0.03cm}
		(b)\includegraphics[width=7.5cm]{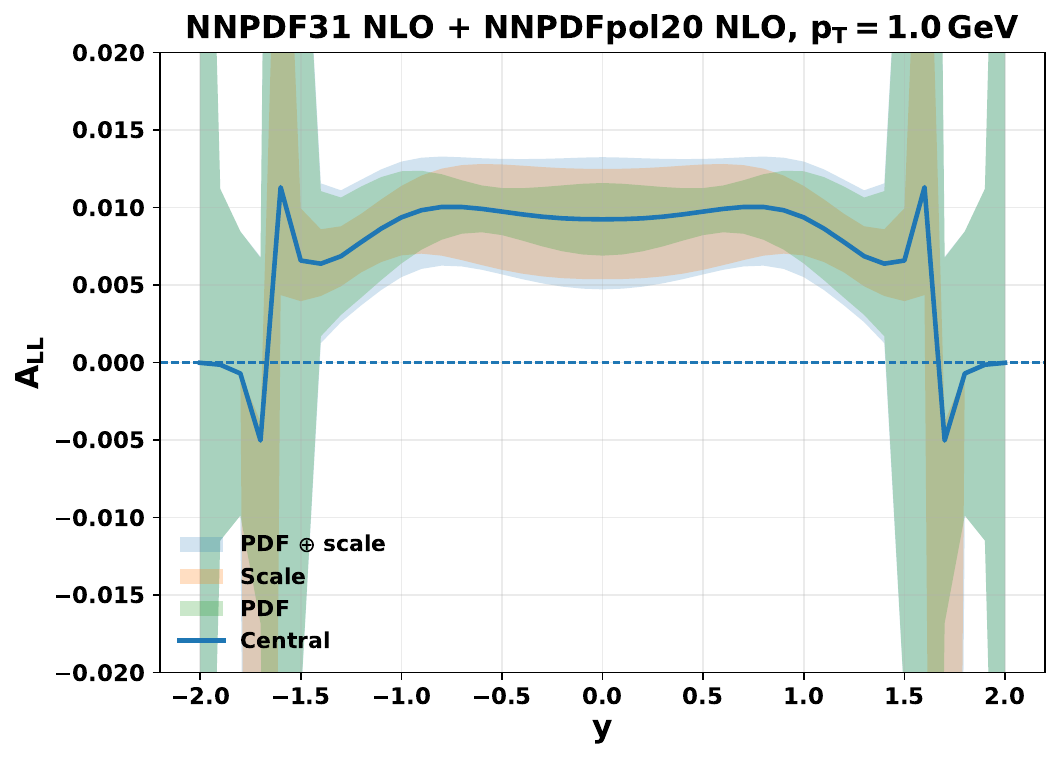}
		\hspace{0.03cm} 
        (c)\includegraphics[width=7.5cm]{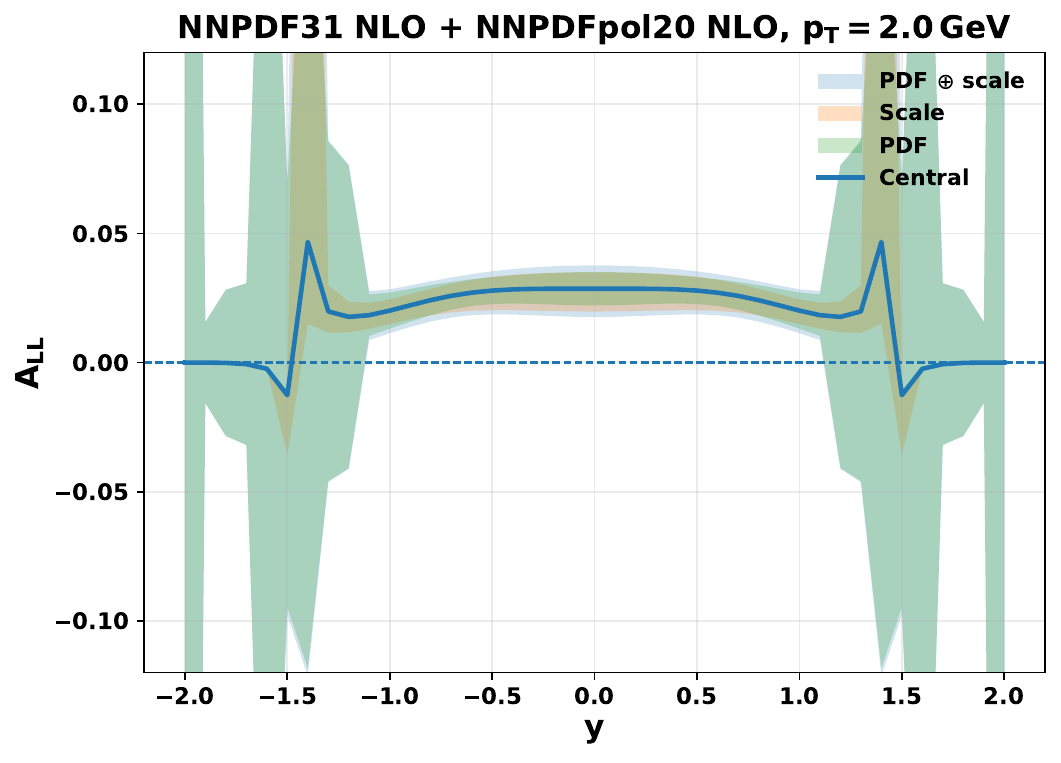}
		\hspace{0.03cm} \\
	\end{minipage}
	\caption{\label{fig:ALL_y_uncNLO}  
Longitudinal double-spin asymmetry \(A_{LL}\) for color-singlet \(J/\psi\) production through \(g g\to J/\psi+g\) at \(\sqrt{s}=27\) GeV as a function of \(y_{J/\psi}\) at (a) \(p_T=0.5\) GeV, (b) \(p_T=1.0\) GeV, and (c) \(p_T=2.0\) GeV. The central prediction is obtained with the NNPDF31 NLO and NNPDFpol20 NLO PDF sets. The displayed bands represent the PDF uncertainty, the uncertainty from factorization- and renormalization-scale variations, and the total uncertainty obtained by combining the two sources in quadrature.
}
\end{figure*} 
\subsection{PDF and scale uncertainties}
\label{subsec:uncertainties}

Having established the sensitivity of the longitudinal double-spin
asymmetry to the PDF input, we next quantify the theoretical
uncertainties associated with the PDF parametrization and the
factorization and renormalization scales. In the following, the PDF
and scale uncertainties are evaluated separately, together with their
quadratic combination, for both the LO and NLO PDF configurations.
The corresponding results are shown in
Figs.~\ref{fig:ALL_pt_unc}-\ref{fig:ALL_y_uncNLO}.

At LO, the uncertainty pattern is strongly dependent on the
kinematic region.  For $y_{J/\psi}=0$, the PDF uncertainty becomes
increasingly important with increasing transverse momentum, while the
scale uncertainty remains comparatively moderate over most of the
considered range. Consequently, the PDF contribution provides the
dominant component of the total uncertainty in the high-$p_T$ region.
At $y_{J/\psi}=1$, the same qualitative behavior is observed, although
the PDF uncertainty becomes substantially broader at large $p_T$.
This reflects the increasing sensitivity to the less-constrained
gluon distributions in the moderate-to-large-$x$ region probed at
$\sqrt{s}=27~\mathrm{GeV}$.  In the rapidity distributions, the PDF
uncertainty also increases towards the edges of the considered
rapidity interval, whereas around central rapidity the scale
uncertainty can constitute a non-negligible fraction of the total
uncertainty.

The corresponding NLO results exhibit a somewhat different
uncertainty pattern.  In the central-rapidity region, the NLO PDF
uncertainty is generally reduced compared with the LO result over a
substantial part of the considered $p_T$ range.  The scale uncertainty
also remains relevant, and the relative importance of the two
contributions depends on $p_T$.  At larger transverse momentum, the
PDF uncertainty again becomes increasingly important, although it
should not be regarded as universally dominant over the entire
kinematic range.  The comparison between the LO and NLO results
therefore demonstrates that the smaller NLO PDF uncertainty in the
central region does not translate into a uniform reduction of the
total theoretical uncertainty.

A more pronounced effect is observed for forward rapidity.  For
$y_{J/\psi}=1$, the NLO uncertainty bands become substantially broader
at large $p_T$, and the PDF and scale contributions can both become
large.  In this region, the relative dominance of one uncertainty
source over the other is therefore not robust.  Similarly, for the
rapidity dependence at fixed $p_T$, the NLO PDF uncertainty remains
moderate around central rapidity but increases strongly towards the
edges of the rapidity interval.  The rapid growth of the uncertainty
bands in these regions indicates that the prediction becomes
increasingly sensitive to the poorly constrained partonic momentum
fractions and to the perturbative scale choice.

Overall, the comparison of the LO and NLO uncertainty estimates shows
that the PDF contribution is an important, and often dominant, source
of uncertainty, particularly at large $p_T$ and away from central
rapidity.  However, its dominance is not universal: the scale
uncertainty can be comparable to, or in some regions larger than, the
PDF uncertainty.  Moreover, the NLO PDF uncertainty is generally
smaller than the LO one in the central kinematic region, while this
reduction is not maintained uniformly at large $p_T$ or forward
rapidity.  These observations emphasize the importance of considering
both PDF and scale uncertainties when assessing the theoretical
precision of $A_{LL}$ at SPD energies.


\subsection{Physics implications}
\label{subsec:physics_implications}

Taken together, the results establish several important features of
the longitudinal double-spin asymmetry for color-singlet $J/\psi$
production at SPD energies. First, the asymmetry is positive over the
kinematically stable region of the calculation and increases
substantially with transverse momentum. This enhancement reflects the
increasing importance of larger gluon momentum fractions as $p_T$ is
raised at fixed $\sqrt{s}$. Second, the magnitude of $A_{LL}$ exhibits
a pronounced sensitivity to the choice of the unpolarized PDF,
particularly for $p_T\gtrsim2$-$3~\mathrm{GeV}$. The difference between
the NNPDF31 and CT18 predictions is therefore much larger than the
small PDF-set dependence observed at the lowest transverse momenta.
Third, the uncertainty analysis demonstrates that both the PDF and
factorization- and renormalization-scale uncertainties can be relevant,
with their relative importance depending on the kinematic region.

The simultaneous dependence on $p_T$ and $y_{J/\psi}$ is especially
useful from the phenomenological point of view. Central rapidities
primarily probe more symmetric momentum fractions, whereas increasing
$|y_{J/\psi}|$ generates increasingly asymmetric gluon momentum
fractions. The resulting variation of $A_{LL}$ therefore provides
complementary sensitivity to different regions of the gluon
polarization. In particular, the enhancement of the asymmetry with
$p_T$ makes the moderate-$p_T$ region potentially valuable for
constraining the polarized gluon distribution at relatively large
$x$, while the rapidity dependence provides additional leverage on
different combinations of the momentum fractions of the two incoming
gluons.

The comparison of the LO and NLO PDF inputs provides an additional
perspective on the theoretical uncertainty. In the central kinematic
region, the NLO PDF uncertainty is generally smaller than the
corresponding LO PDF uncertainty over a substantial part of the
considered range. This reduction, however, is not uniform. At larger
$p_T$ and away from central rapidity, the NLO uncertainty bands can
become substantially broader, and the PDF and scale contributions may
both be significant. Consequently, the PDF uncertainty should not be
regarded as universally dominant over the entire phase space. Rather,
the relative importance of the PDF and scale uncertainties is
kinematics dependent, with PDF uncertainties becoming particularly
important in regions probing less-constrained polarized gluon
distributions.

This behavior has direct implications for the interpretation of a
future measurement. In regions where the PDF uncertainty is dominant,
the theoretical prediction is particularly sensitive to the polarized
gluon distribution and a precise measurement of $A_{LL}$ could provide
useful additional information on gluon helicity. Conversely, in
regions where the scale uncertainty is comparable to the PDF
uncertainty, improvements in the perturbative description of the hard
process will also be important for extracting quantitative information
from the measurement. The comparison between LO and NLO PDF inputs
therefore indicates that both the quality of the polarized PDFs and
the perturbative accuracy of the hard-scattering calculation are
relevant for exploiting the constraining power of $A_{LL}$.

It is important, however, to distinguish these robust phenomenological
features from numerical structures appearing close to the boundaries
of phase space. Several PDF combinations exhibit sharp excursions
when $|y_{J/\psi}|$ approaches its kinematic limit, while the
NNPDF31-NLO combination develops an isolated, very large structure in
the forward-rapidity $p_T$ distribution. Since $A_{LL}$ is a ratio of
polarized and unpolarized cross sections, such structures can be
strongly amplified when the denominator becomes small or poorly
determined. They should therefore not be interpreted as evidence for a
physical enhancement or sign change of the asymmetry. The
phenomenologically stable central region, where the different PDF
replicas and the denominator remain well behaved, provides the
appropriate basis for quantitative conclusions.

Finally, the present calculation should be interpreted as a
leading-order hard-scattering study with alternative PDF orders rather
than as a complete NLO prediction. The comparison of LO and NLO PDF
inputs is nevertheless useful because it exposes the sensitivity of
the observable to the perturbative order of the PDF evolution and to
differences between global PDF analyses. A complete NLO calculation
of the hard coefficient would be required before attributing the
observed LO-NLO-PDF differences to genuine higher-order corrections.
Within the present accuracy, the strong PDF dependence and the
increasing uncertainties at larger $p_T$ identify the kinematic
regions in which future polarized $J/\psi$ measurements at SPD could
provide additional information on the gluon helicity distribution.
%
\section{Conclusion}
\label{sec:conclusion}

We have presented a phenomenological study of the longitudinal
double-spin asymmetry $A_{LL}$ in color-singlet $J/\psi$ production
through the gluon-fusion subprocess
$g g\to J/\psi+g$ in longitudinally polarized proton-proton
collisions at $\sqrt{s}=27$~GeV, corresponding to the upper energy
anticipated for the SPD at the NICA collider.
The calculation is performed at leading order for the hard-scattering
coefficient, while several combinations of polarized and unpolarized
PDFs are employed to investigate the sensitivity of the observable to
the gluon helicity distributions and to the perturbative order of the
PDF input. The phenomenological analysis is supplemented by an
assessment of PDF and factorization- and renormalization-scale
uncertainties.

The calculated asymmetry is found to be positive throughout the
kinematically stable region and exhibits a pronounced dependence on
the transverse momentum. At central rapidity, $A_{LL}$ increases from
values of order $10^{-3}$ at low $p_T$ to values approaching $0.2$ at
$p_T=5$~GeV for the NNPDF31 LO-NNPDFpol20 LO combination. The increase
with $p_T$ reflects the increasing sensitivity of the process to
moderate-to-large gluon momentum fractions accessible at the
relatively low center-of-mass energy of SPD. The rapidity dependence
provides complementary information by probing increasingly asymmetric
momentum fractions as $|y_{J/\psi}|$ increases. The symmetry of the
calculated distributions under $y_{J/\psi}\to-y_{J/\psi}$ is
consistent with the identical polarized proton beams.

A significant dependence on the choice of PDF set is observed,
particularly at moderate and large transverse momentum. The NNPDF31
and CT18 based predictions can differ substantially, with the
difference becoming increasingly pronounced as the process probes
larger gluon momentum fractions. This demonstrates that $J/\psi$
production at SPD energies can provide sensitivity to features of the
polarized gluon distribution that are not readily distinguished in the
low-$p_T$ region. The comparison between LO and NLO PDF inputs further
shows that the numerical prediction for $A_{LL}$ can be appreciably
affected by the perturbative order of the input PDFs. Since the
hard-scattering coefficient is kept at LO, however, these comparisons
should be regarded as a study of PDF-order dependence and not as a
complete NLO prediction.

The uncertainty analysis provides an important qualification to this
sensitivity. The relative importance of PDF and scale uncertainties
depends on the kinematic region. In the central region, the NLO PDF
uncertainty is generally smaller than the corresponding LO PDF
uncertainty over a substantial part of the considered range, while
the scale uncertainty remains non-negligible. At larger $p_T$ and
away from central rapidity, the PDF uncertainty can increase
substantially and can become the dominant contribution, but there are regions in which the PDF and scale uncertainties are comparable.
Thus, the theoretical uncertainty of $A_{LL}$ cannot be characterized
by a universal hierarchy between PDF and scale variations. In
particular, the broad uncertainty bands observed at large $p_T$ and
forward rapidity emphasize the need for improved polarized PDFs and,
for precision phenomenology, a more complete treatment of higher-order
hard-scattering corrections.

We also find that the ratio defining $A_{LL}$ becomes numerically
unstable close to the kinematic boundaries, where the unpolarized
cross section becomes small. Sharp endpoint structures in the
rapidity distributions, as well as isolated excursions for particular
PDF combinations in the high-$p_T$, forward-rapidity region, are
therefore not regarded as robust physical predictions. The
phenomenologically reliable information is provided by the region in
which the polarized and unpolarized cross sections remain numerically
stable and the PDF ensembles give well-defined ratios.

Overall, our results indicate that longitudinally polarized
$J/\psi$ production at SPD energies offers a promising probe of the
polarized gluon content of the proton in the moderate-to-large-$x$
region. The strong $p_T$ and PDF dependence of $A_{LL}$, together with
its sensitivity to the choice and perturbative order of the polarized
gluon input, makes this observable complementary to existing
constraints from polarized deep-inelastic scattering and other
spin-dependent processes. A future measurement of the asymmetry at
NICA, particularly with differential information in both $p_T$ and
$y_{J/\psi}$, could therefore provide valuable additional constraints
on the gluon helicity distribution. A complete NLO calculation of the
hard-scattering coefficient, together with a systematic treatment of
higher order corrections and the corresponding PDF uncertainties,
will be important for making precision predictions and for fully
exploiting the constraining power of such measurements.
%
%
%
\section*{Acknowledgements}
This work was done with support from Ministry of Science and Higher Education of Russian Federation under State Assignment № 075-03-2025-662 from 17.01.2025. The authors are grateful to Prof. W.~C.~Chang for providing the data from his article \cite{Hsieh:2021yzg}, which were used to plot Fig.~\ref{contrilog}.
\section*{Appendix}
%
%
\begin{table}[]
\centering
\caption{2 $\to$ 2 Kinematics}
{\scriptsize\begin{tabular}{|c|c|c|c|c|}
\hline
$p_T$ [GeV] & $y$ & $y_g$ & $x_1$ & $x_2$ \\
\hline
0.50 & 0.00 & 0.00 & 0.13 & 0.13 \\
0.50 & 0.00 & 1.00 & 0.16 & 0.12 \\
0.50 & 0.00 & 1.50 & 0.20 & 0.12 \\
0.50 & 0.00 & 2.00 & 0.25 & 0.12 \\
0.50 & 1.00 & 0.00 & 0.33 & 0.06 \\
0.50 & 1.00 & 1.00 & 0.36 & 0.05 \\
0.50 & 1.00 & 1.50 & 0.39 & 0.05 \\
0.50 & 1.00 & 2.00 & 0.44 & 0.04 \\
0.50 & 1.50 & 0.00 & 0.52 & 0.04 \\
0.50 & 1.50 & 1.00 & 0.56 & 0.03 \\
0.50 & 1.50 & 1.50 & 0.59 & 0.03 \\
0.50 & 1.50 & 2.00 & 0.64 & 0.03 \\
0.50 & 2.00 & 0.00 & 0.85 & 0.03 \\
0.50 & 2.00 & 1.00 & 0.88 & 0.02 \\
0.50 & 2.00 & 1.50 & 0.92 & 0.02 \\
0.50 & 2.00 & 2.00 & 0.97 & 0.02 \\
1.50 & 0.00 & 0.00 & 0.18 & 0.18 \\
1.50 & 0.00 & 1.00 & 0.28 & 0.15 \\
1.50 & 0.00 & 1.50 & 0.37 & 0.14 \\
1.50 & 0.00 & 2.00 & 0.54 & 0.13 \\
1.50 & 1.00 & 0.00 & 0.39 & 0.10 \\
1.50 & 1.00 & 1.00 & 0.49 & 0.07 \\
1.50 & 1.00 & 1.50 & 0.59 & 0.06 \\
1.50 & 1.00 & 2.00 & 0.75 & 0.05 \\
1.50 & 1.50 & 0.00 & 0.61 & 0.08 \\
1.50 & 1.50 & 1.00 & 0.71 & 0.05 \\
1.50 & 1.50 & 1.50 & 0.81 & 0.04 \\
1.50 & 1.50 & 2.00 & 0.97 & 0.04 \\
1.50 & 2.00 & 0.00 & 0.97 & 0.07 \\
1.50 & 2.00 & 1.00 & \multicolumn{1}{|c}{--} & \multicolumn{1}{|c|}{--} \\
1.50 & 2.00 & 1.50 & \multicolumn{1}{|c}{--} & \multicolumn{1}{|c|}{--} \\
1.50 & 2.00 & 2.00 & \multicolumn{1}{|c}{--} & \multicolumn{1}{|c|}{--} \\
3.00 & 0.00 & 0.00 & 0.27 & 0.27 \\
3.00 & 0.00 & 1.00 & 0.46 & 0.20 \\
3.00 & 0.00 & 1.50 & 0.66 & 0.18 \\
3.00 & 0.00 & 2.00 & 0.98 & 0.17 \\
3.00 & 1.00 & 0.00 & 0.54 & 0.17 \\
3.00 & 1.00 & 1.00 & 0.73 & 0.10 \\
3.00 & 1.00 & 1.50 & 0.93 & 0.08 \\
3.00 & 1.00 & 2.00 & \multicolumn{1}{|c}{--} & \multicolumn{1}{|c|}{--} \\
3.00 & 1.50 & 0.00 & 0.82 & 0.15 \\
3.00 & 1.50 & 1.00 & \multicolumn{1}{|c}{--} & \multicolumn{1}{|c|}{--} \\
3.00 & 1.50 & 1.50 & \multicolumn{1}{|c}{--} & \multicolumn{1}{|c|}{--} \\
3.00 & 1.50 & 2.00 & \multicolumn{1}{|c}{--} & \multicolumn{1}{|c|}{--} \\
3.00 & 2.00 & 0.00 & \multicolumn{1}{|c}{--} & \multicolumn{1}{|c|}{--} \\
3.00 & 2.00 & 1.00 & \multicolumn{1}{|c}{--} & \multicolumn{1}{|c|}{--} \\
3.00 & 2.00 & 1.50 & \multicolumn{1}{c}{--} & \multicolumn{1}{|c|}{--} \\
3.00 & 2.00 & 2.00 & \multicolumn{1}{c}{--} & \multicolumn{1}{|c|}{--} \\
5.00 & 0.00 & 0.00 & 0.40 & 0.40 \\
5.00 & 0.00 & 1.00 & 0.72 & 0.28 \\
5.00 & 0.00 & 1.50 & \multicolumn{1}{c}{--} & \multicolumn{1}{|c|}{--} \\
5.00 & 0.00 & 2.00 & \multicolumn{1}{c}{--} & \multicolumn{1}{|c|}{--} \\
5.00 & 1.00 & 0.00 & 0.77 & 0.27 \\
5.00 & 1.00 & 1.00 & \multicolumn{1}{c}{--} & \multicolumn{1}{|c|}{--} \\
5.00 & 1.00 & 1.50 & \multicolumn{1}{c}{--} & \multicolumn{1}{|c|}{--} \\
5.00 & 1.00 & 2.00 & \multicolumn{1}{c}{--} & \multicolumn{1}{|c|}{--} \\
5.00 & 1.50 & 0.00 & \multicolumn{1}{c}{--} & \multicolumn{1}{|c|}{--} \\
5.00 & 1.50 & 1.00 & \multicolumn{1}{c}{--} & \multicolumn{1}{|c|}{--} \\
5.00 & 1.50 & 1.50 & \multicolumn{1}{c}{--} & \multicolumn{1}{|c|}{--} \\
5.00 & 1.50 & 2.00 & \multicolumn{1}{c}{--} & \multicolumn{1}{|c|}{--} \\
5.00 & 2.00 & 0.00 & \multicolumn{1}{c}{--} & \multicolumn{1}{|c|}{--} \\
5.00 & 2.00 & 1.00 & \multicolumn{1}{c}{--} & \multicolumn{1}{|c|}{--} \\
5.00 & 2.00 & 1.50 & \multicolumn{1}{c}{--} & \multicolumn{1}{|c|}{--} \\
5.00 & 2.00 & 2.00 & \multicolumn{1}{c}{--} & \multicolumn{1}{|c|}{--} \\
\hline
\end{tabular}
\label{table2to2}
}
\end{table}
%

\bibliography{ref_SS.bib}

%
\end{document}